\documentclass[article,10pt]{memoir}
\usepackage{microtype}
\usepackage{amsmath, amsthm, amssymb}
\usepackage[overload]{textcase}
\usepackage[round]{natbib}
\usepackage{pdfpages}
\usepackage[colorlinks=true,citecolor=blue,linkcolor=blue,urlcolor=blue]{hyperref}
\usepackage[utf8]{inputenc}
\usepackage{enumitem}

\usepackage{empheq}
\usepackage{xfrac}
\usepackage{multirow}

\usepackage[osf,sc]{mathpazo}
\usepackage[margin=1in]{geometry}

\definecolor{light_blue}{rgb}{0.8, 0.888151, 1.}
\colorlet{shadecolor}{light_blue}
\newcommand*\mybluebox[1]{ 
\colorbox{shadecolor}{\hspace{1em}#1\hspace{1em}}}

\newcommand{\ens}[1]{\left\langle#1\right\rangle}
\newcommand{\xav}{\bar x}
\newcommand{\mpd}{\text{MPD}}

\newcommand{\vstrutSS}{\vphantom{\sum_X^X}}
\newcommand{\kmax}{k_\text{max}}
\newcommand{\sol}{\text{sol}}
\newcommand{\gel}{\text{gel}}

\pretitle{\centering\sffamily\huge}
\posttitle{\vskip 0.5em}
\maxtocdepth{section}   
\maxsecnumdepth{section} 
\chapterstyle{article}

\setsecheadstyle{\large\bfseries\raggedright} 
\setsubsecheadstyle{\normalsize\bfseries\raggedright}

\title{The Smoluchowski Ensemble: Statistical Mechanics of Aggregation}
\author{
 Themis Matsoukas\\
 Department of Chemical Engineering\\
 Penn State University\\
 ~\\
 Entropy 2020, 22(10), 1181; https://doi.org/10.3390/e22101181
}

\begin{document}
\maketitle
\thispagestyle{empty}
\begin{abstract}
We present a rigorous thermodynamic treatment of irreversible binary aggregation.
We construct the Smoluchowski ensemble as the set of discrete finite distributions that are reached in fixed number of merging events and define a probability measure on this ensemble, such that the mean distribution in the mean-field approximation is governed by the Smoluchowski equation. In the scaling limit this ensemble gives rise to a set of relationships identical to those of familiar statistical thermodynamics. 
The central element of the thermodynamic treatment is the selection functional, a functional of feasible distributions that connects the probability of distribution to the details of the aggregation model. 
We obtain scaling expressions for general kernels and closed-form results for the special case of the constant, sum and product kernel. We study the stability of the most probable distribution, provide criteria for the sol-gel transition and obtain the distribution in the post-gel region by simple thermodynamic arguments. 
\end{abstract}

\tableofcontents
\chapter{Introduction}
Aggregation  is the process of forming structures through the merging of clusters.  This generic process is encountered in a large variety of systems, from polymerization and colloidal aggregation to the clustering of social groups and the merging of galaxies.  The mathematical foundations of aggregation were set by Smoluchowski \citep{Smoluchowski:ZFPC17}, whose particular interest was in Brownian coagulation.  The aggregation equation, more commonly known as Smoluchowski equation, is a rate equation on a distribution of clusters whose size (mass) changes by binary aggregation events. For a  discrete population of clusters with integer masses in multiples of a unit mass (``monomer'') it takes the form \citep{Smoluchowski:ZFPC17},

%
\begin{equation}
\label{smoluchowski:0}
   \frac{d c_k}{d t}
   =
   \frac{1}{2}\sum_{j=1}^{k-1} c_{k-j} c_j K_{k-j,j}
   -\sum_{j=1}^\infty c_k c_j  K_{k,j} ,  
\end{equation}
%

\noindent where $c_k$ is the number concentration of clusters with mass $k$ and $K_{i,j}$ the aggregation kernel, a rate constant for the merging of masses $i$ and $j$.   A large body of literature has focused on the theory of the Smoluchowski equation, the existence of analytic solution and the scaling limit \citep{Leyvraz:PR03}. Of particular interest is  \textit{gelling}, a condition that arises under the product kernel $K_{i,j} = i j$; it  refers to the formation of a giant structure, as in polymer gels, and is manifested by the failure of the Smoluchowski equation to conserve mass. This process is commonly described as a phase transition, suggesting the possibility that statistical thermodynamics, a theory developed for equilibrium states of interacting particles, may perhaps be applicable in this clearly irreversible process. 

Studies of Smoluchowski aggregation broadly fall in one of two categories, kinetic and stochastic. The kinetic approach is based on Eq.\ (\ref{smoluchowski:0}) and its solution. Stable solutions conserve mass; gelling is identified as the point where mass conservation breaks down   \citep{Ziff:PRL82,Hendriks:JSP83}. Post-gel solutions require additional assumptions as to how the gel and the dispersed phase interact \citep{Ziff:JCP80}. 
The stochastic approach views clusters as entities that merge with probabilities proportional to the aggregation kernel.  It was first formulated by Marcus \citep{Marcus:T68} for a discrete finite population, and its formal mathematical treatment was developed by Lushnikov, who obtained solutions for certain special cases,  including gelation \citep{Lushnikov:JCIS78,Lushnikov:JPMT11,Lushnikov:PRE05,Lushnikov:JPMG05a,Lushnikov:PRL04}. In Lushnikov's method all feasible distributions are given a probability, whose evolution in time is tracked via a generating functional.  The approach is explicitly probabilistic and views the Smoluchowski equation as the mean-field approximation of the underlying stochastic process \citep{Aldous:B99}. 
A different approach within the probabilistic realm makes use of combinatorial methods. This treatment originated with  \citet{Stockmayer:JCP43} and was further explored by Spouge 
\citep{Spouge:JPMG85,Spouge:M83_121,Hendriks:ZPCM85B,Spouge:1983}. The combinatorial approach considers the number of ways to build a particular distribution of clusters  and assigns probabilities in proportion to that combinatorial weight. The ensemble of distributions is then reduced to the most probable distribution, which is identified by maximizing the combinatorial weight. This approach has two appealing advantages. It deals with a time-free ensemble in which time appears implicitly via the mean cluster mass. More importantly, it brings the problem closer to the viewpoint of statistical mechanics and the notion that an ensemble may be represented in the scaling limit by its most probable element. Stockmayer recognized this connection and his treatment of gelation is replete with references to the theory of phase transitions \citep{Stockmayer:JCP43}. The analogy between aggregation and thermodynamics was not formalized, however. Stockmayer obtained the gel point by mathematical, not thermodynamic methods, and arrived at a post-gel solution that is not consistent with the kinetics of aggregation \citep{Ziff:JCP80}. 

We have previously shown that gelation can be indeed treated as a formal phase transition  and have presented solutions for the product kernel in the pre- and post-gel regions \citep{Matsoukas:SR2015} based on our earlier work on the cluster ensemble \citep{Matsoukas:PR15,Matsoukas:statPBE_PRE14}. Here we generalize the methodology to formulate a rigorous thermodynamic theory of Smoluchowski aggregation. 
We begin with a finite population that starts from a well defined state and construct the set of all possible distributions that can be reached in a fixed number of elementary transitions.  The probability of distribution in this ensemble is governed by the kinetics of the elementary processes that act on the population. In the thermodynamic limit the most probable distribution is overwhelmingly more probable than all others and is governed by a set of mathematical relationships that we recognize as \textit{thermodynamics}. The work is organized as follows. 
In Section \ref{sct:ensemble} we define the Smoluchowski ensemble of distributions and their probabilities. 
In Section \ref{sct:thermo} we formulate the probability of distribution in terms of a special functional $W$ that introduces the partition function and the Shannon entropy of distribution.
In Section \ref{sct:thl} we treat the scaling limit and derive the thermodynamic relationships of the Smoluchowski ensemble. 
In Section \ref{sct:gibbs} we obtain solutions of the Gibbs form for the classical kernels, constant, sum and product. 
We analyze the stability and phase behavior of the ensemble in Section \ref{sct:stability} and treat the sol-gel process as a phase transition. 
In Section \ref{sct:cont} we express the results in the continuous domain and finally offer concluding remarks in Section \ref{sct:summary}.

\chapter{The Smoluchowski Ensemble}\label{sct:ensemble}
We consider a population of clusters composed of $i = 1,2\cdots$ units (monomers). In binary aggregation two clusters merge to form a new cluster that conserves mass, via the schematic reaction

%
\begin{equation}\label{rxn}
   (i)+(j) \xrightarrow{K_{i,j}} (i+j) .
\end{equation}
%

\noindent The merging of a pair constitutes an elementary stochastic event, whose probability depends on the aggregation kernel $K_{i,j}$.  At the initial state the population consists of $N_0 = M$ single members (monomers). This distribution constitutes generation $g=0$. The next generation is constructed by implementing every possible aggregation event in the distribution of generation $g=0$. The set of distributions formed in this manner constitutes the microcanonical ensemble of generation $g=1$. We continue recursively to form the ensemble of distributions in generation $g$ by implementing all possible aggregation events, one at a time, in all distributions of the parent ensemble. We represent a distribution of clusters by the vector $\mathbf{n}=(n_1,n_2\cdots)$, where $n_i$ is the number of clusters with $i$ members. All distributions in generation $g$ satisfy the conditions

%
\begin{equation}\label{constraints}
   \sum_i n_i = M-g = N,\quad
   \sum_i i n_i = M .
\end{equation}
%

\noindent The first condition expresses the fact each elementary event decreases the number of clusters by 1, according to the stoichiometry of binary merging; the second condition expresses the fact that the number of members is conserved. Conversely, any distribution that satisfies the conditions in Eq.\ (\ref{constraints}) is a member of the ensemble of generation $g$ because it can be formed in $g$ steps from $M$ monomers. We view the two equations in Eq.\ (\ref{constraints}) as the constraints that define the ensemble of feasible distributions. We call this ensemble microcanonical to indicate that it is conditioned by two extensive constraints that fix the mean cluster mass $M/N=\xav$ in all distributions of the ensemble. 

\begin{figure}
\begin{center}
\includegraphics[width=3.25in]{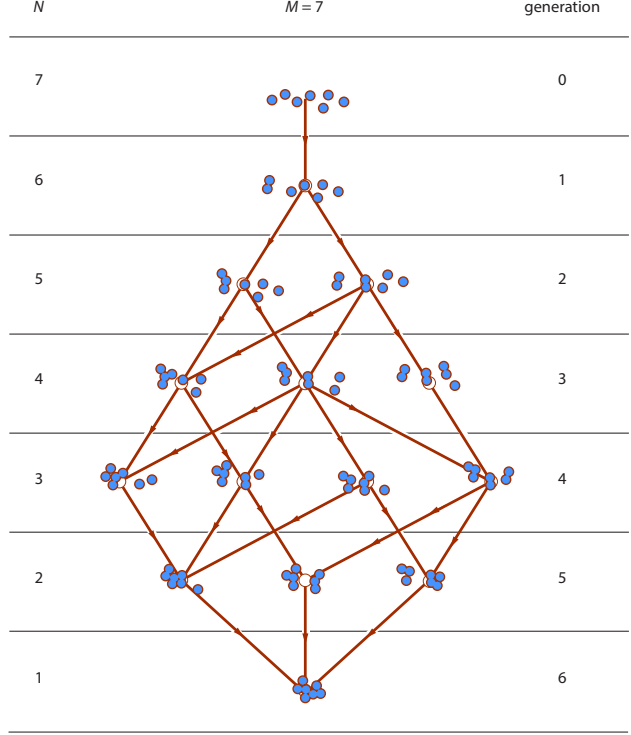}
\end{center}
\caption{The aggregation graph for $M=7$. Each layer contains all feasible distributions in that generation. }
\label{fig1}
\end{figure}

The evolution of the ensemble may be represented in the form of a layered graph (Fig.\ \ref{fig1}), whose vertices represent distributions and edges represent elementary transitions according to Eq.\ (\ref{rxn}). 
Edges are directed from parent in generation $g-1$ to offspring in generation $g$. Layers are organized by generation and contain all distributions in a generation. The graph begins in generation $g=0$ with a distribution of all monomers and ends when  all units have joined the same cluster. 
Stochastic aggregation is a random walk on this graph. A trajectory is a possible sequence of connected edges from top to bottom. Our goal is to establish the probability $P(\mathbf{n})$ of distribution in generation $g=0,1,\cdots M-1$,  in terms the aggregation kernel $K_{i,j}$ for any $M$.  

\section{Kinetics}
When cluster masses $i-j$ and $j$, in distribution $\mathbf{n'}$ of generation $g-1$, merge to form a cluster of mass $i$,  parent distribution $\mathbf{n'}$ is transformed to offspring distribution $\mathbf{n}$ via the transition

%
\begin{equation}\label{transition:rxn}
   \mathbf{n'} \xrightarrow{(i-j)+(j)\to (i) }\mathbf{n} .
\end{equation}
%

\noindent This transition is represented by an edge in the graph of Fig.\ \ref{fig1}. Its rate $R_{i-j,j}$ is proportional to the number of ways to choose the reactants and the proportionality factor is the aggregation kernel: 

%
\begin{equation}   
\label{transition_rate}
   R_{i-j,j} = K_{i-j,j} \frac{n'_{i-j} (n'_j-\delta_{i-j,j})}{1+\delta_{i-j,j}} . 
\end{equation}
%

\noindent The total rate by which parent $\mathbf{n'}$ produces offspring is

%
\begin{equation}   R(\mathbf{n'}) = K(\mathbf{n'}) \frac{N'(N'-1)}{2} , 
\end{equation}
%

\noindent where $N' = \sum_i n'_i$ is the number of clusters and $K(\mathbf{n'})$ is the mean kernel in parent distribution $\mathbf{n'}$:

%
\begin{equation}   K(\mathbf{n'}) = 
   \frac{2}{N'(N'-1)}\sum_i\sum_j K_{i,j} 
   \frac{n'_i (n'_j-\delta_{i,j})}{1+\delta_{i,j}} . 
\end{equation}
%

In physical terms the aggregation kernel $K_{i,j}$ is the rate constant for the reaction between masses $i$ and $j$. Its mathematical form may be constructed on the basis of a kinetic model for the particular problem. It is beyond the scope of this work to review the numerous kernels that have been proposed in the literature. We mention a selected few that are important for their physical, mathematical and historical significance, and summarize them in  Table \ref{tbl_Kij}. 

The Brownian coagulation kernel was derived by \citet{Smoluchowski:ZFPC17} to describe the kinetics of diffusion limited aggregation in colloidal systems. 
The constant kernel was adopted by \citet{Smoluchowski:ZFPC17} as an approximation for the Brownian kernel, a simplification that allows analytic results. This kernel is obtained by setting $i=j$ in the Brownian kernel. 
The Flory/Stockmayer kernel \citep{Flory:JACS41a,Stockmayer:JCP43} is a model for polymerization of chains composed of monomers with $f$ functional groups. Assuming no cycles, a polymer with $i$ monomers contains $f i - 2 i + 2$ unreacted functional groups that are available to react. The Flory/Stockmayer kernel is the product of the unreacted functional groups in the two chains that merge. This kernel leads to gelation \citep{Stockmayer:JCP43}.
The product kernel is the limiting form of the Flory/Stockmayer kernel when the number of functional groups approaches infinity. It also leads to gelation, and being a simpler kernel than the Flory/Stockmayer, it serves as the standard model to study gelation. 
The sum kernel is proportional to the number of units in each cluster. This kernel may be viewed as the limiting form of a Flory/Stockmayer type kernel with two kinds functional groups \citep{Spouge:M83_121}, but its significance is primarily mathematical as one of a handful of kernels that lead to analytic solutions. 

We discuss the constant, sum and product kernel in detail in Section \ref{sct:gibbs}. For now we leave the kernel general and unspecified. We only place the minimum conditions, $K_{i,j}=K_{j,i}>0$, which are required from elementary physical considerations; additionally, we adopt the normalization $K_{1,1}=1$.

\begin{table}
\caption{Selected aggregation kernels}
\label{tbl_Kij}
\begin{equation*}
\renewcommand{\arraystretch}{2}
\begin{array}{ll}
\toprule
   \text{Brownian coagulation} 
   & \displaystyle 
     K_{i,j} = \frac{1}{4}\left(
     2
     +\left(\frac{i}{j}\right)^{1/3}
     +\left(\frac{j}{i}\right)^{1/3}
\right)
\\
   \text{Constant kernel}   
   & K_{i,j} = 1 
\\
   \text{Flory/Stockmayer kernel} \quad~
   & \displaystyle 
     K_{i,j} = \frac{(f i - 2 i + 2)(f j - 2 j + 2)}{f^2}
\\
   \text{Product kernel}
   & K_{i,j} = i j
\\
   \text{Sum kernel} 
   & \displaystyle 
     K_{i,j} = \frac{i+j}{2}
     \rule[-15pt]{0pt}{0pt}
\\
\toprule
\end{array} 
\end{equation*}
\end{table}

\section{Probabilities}\label{sct:probabilities}
We assign a probability $P(\mathbf{n})$ to each distribution $\mathbf{n}$ within generation $g$ and formulate the propagation of probability between generations as follows: 

%
\begin{equation}\label{master}
   P(\mathbf{n}) = \sum_\mathbf{n'} P(\mathbf{n'})  \frac{R_{i-j,j}}{\ens{R}_{g-1}} .
\end{equation}
%

\noindent Here $\mathbf{n}$ is a distribution in generation $g$, $\mathbf{n'}$ is its parent of $\mathbf{n}$ in generation $g-1$ via the reaction $(i-j)+(j)\to (i)$, $R_{i-j,j}$ is the rate of the reaction, and $\ens{R}_{g-1}$ is the mean reaction rate in parent generation $g-1$:

%
\begin{equation}\label{ens:Rg-1}
   \ens{R}_{g-1} = \sum_\mathbf{n'} P(\mathbf{n'}) R(\mathbf{n'}) . 
\end{equation}
%

\noindent In both Eq.\ (\ref{master}) and (\ref{ens:Rg-1}) the summations are over all distributions $\mathbf{n'}$ in generation $g-1$.  Expressing the transition rate in terms of the aggregation kernel we obtain

%
\begin{equation}\label{T:def}
   \frac{R_{i-j,j}}{\ens{R}_{g-1}} 
   = 
   \frac{2}{N'(N'-1)} 
   \frac{K_{i-j,j}}{\ens{K}_{g-1}}
   \frac{n'_{i-j} (n'_j-\delta_{i-j,j})}{1+\delta_{i-j,j}} .
\end{equation}
%

\noindent with

%
\begin{equation}   \ens{K}_{g-1} = \sum_\mathbf{n'} P(\mathbf{n'}) K(\mathbf{n'}) .
\end{equation}
%

\noindent We may confirm that $P(\mathbf{n})$ as defined in Eq.\ (\ref{master}) satisfies normalization over all distributions in the same generation. Beginning with $P(\mathbf{n}_0)=1$ at the initial state, Eq.\ (\ref{master}) uniquely determines the probabilities of all distributions in all future generations.

\section{Smoluchowski Equation}
The mean  number of clusters with mass $k$ in generation $g$ is

\begin{equation}\label{menad:dstr:def}
   \ens{n_k} = \sum_\mathbf{n} n_k P(\mathbf{n}) ,
\end{equation}

\noindent with the summation going over all distributions in the same generation. We will derive the evolution of the mean distribution from parent generation $g-1$ to generation $g$. 
The probability of distribution $P(\mathbf{n})$ is given by Eq.\ (\ref{master}) and is expressed as a summation over its parents $\mathbf{n'}$.  By the stoichiometry of the transition in Eq.\ (\ref{rxn}),  the parent and offspring distributions satisfy

%
\begin{equation}\label{stoichio}
   n_k = n'_k+\delta_{k,i} - \delta_{k,i-j} - \delta_{k,j} . 
\end{equation}
%

\noindent Combining this relationship with (\ref{menad:dstr:def}) and (\ref{master}) the result is (see Supplementary Material)

\begin{multline}
\label{smoluchowski_1}
   \ens{n_k} - \ens{n'_k} 
   =
   \frac{2}{N(N+1)\ens{K}_{M,N+1}}  
   \left\langle
   \frac{1}{2}\sum_{j=1}^{k-1} n_{k-j}(n_j-\delta_{k-j,j})K_{k-j,j}
   -\sum_{j=1}^\infty n_k (n_j-\delta_{k,j}) K_{k,j}
   \right\rangle_{M,N+1} .
\end{multline}

\noindent The left-hand side is the change in the mean number of $k$-mers between generations; the right-hand side is the ensemble average of the production and depletion of $k$-mers within all distributions of the parent ensemble. Define the mean time increment $\Delta t$ from parent to offspring generation as

%
\begin{equation}   \Delta t\Big|_{g-1\to g} = \frac{2}{N(N+1) \ens{K}_{M,N+1}}  ;
\end{equation}
%

\noindent then Eq.\ (\ref{smoluchowski_1}) reads

%
\begin{equation}\label{smoluchowski_exact}
   \left.\frac{\Delta \ens{n_k}}{\Delta t}\right|_{g-1\to g}
   =
   \left\langle
   \frac{1}{2}\sum_{j=1}^{k-1} n_{k-j}(n_j-\delta_{k-j,j})K_{k-j,j}
   -\sum_{j=1}^\infty n_k (n_j-\delta_{k,j}) K_{k,j}
   \right\rangle_{M,N+1} .
\end{equation}
%

\noindent In the mean-field approximation we reduce the ensemble into a single distribution, $\mathbf{n^*}$. This resolves the ensemble averages trivially and leads to the governing equation for $\mathbf{n^*}$:

\begin{empheq}[box=\mybluebox]{equation}
\label{smoluchowski}
   \frac{\Delta n^*_k}{\Delta t}
   =
   \frac{1}{2}\sum_{j=1}^{k-1} n^*_{k-j}(n^*_j-\delta_{k-j,j})K_{k-j,j}
   -\sum_{j=1}^\infty n^*_k (n^*_j-\delta_{k,j}) K_{k,j} . 
\end{empheq}

\noindent This is the Smoluchowski equation for binary aggregation, the discrete finite equivalent of Eq.\ (\ref{smoluchowski:0}).  The mean field approximation, which is invoked to obtain (\ref{smoluchowski}), implies that a single distribution is representative of the entire ensemble. In Sections \ref{sct:thl} and \ref{sct:stability} we examine the conditions under which this is true. 

\chapter{Thermodynamic Formalism}\label{sct:thermo}
\section{Partition Function and Selection Functional}
We now formulate the probability of distribution in terms of a special functional, $W(\mathbf{n})$. It is through this formulation that we will make contact with statistical thermodynamics. 
We begin by writing the probability $P(\mathbf{n})$ in generation $g$ in the form

\begin{empheq}[box=\mybluebox]{equation}
\label{Prob:n}
   P(\mathbf{n}) = \frac{\mathbf{n!}W(\mathbf{n})}{\Omega_{M,N}} ,
\end{empheq}

\noindent where $\mathbf{n!}$ is the multinomial coefficient of vector $\mathbf{n}$,

%
\begin{equation}   
    \mathbf{n!} 
       = \frac{(n_1+n_2\cdots)!}{n_1! n_2! \cdots} 
       = \frac{N!}{n_1! n_2! \cdots} , 
\end{equation}
%

\noindent $N=M-g$ is the number of clusters in all distributions of generation $g$,  $W(\mathbf{n})$ is a functional of distribution $\mathbf{n}$, to be determined, and $\Omega_{M,N}$ is the partition function. By the normalization condition on $P(\mathbf{n})$ the partition function satisfies
%
\begin{equation}\label{Omega:def}
   \Omega_{M,N} = \sum_\mathbf{n} \mathbf{n!} W(\mathbf{n}),
\end{equation}
%

\noindent with the summation over all distributions in generation $g=M-N$.  

\section{Shannon Entropy}
The multinomial coefficient represents the combinatorial multiplicity of distribution $\mathbf{n}$, namely, the number of ways to order the clusters in the distribution, if clusters with the same number of units are treated as indistinguishable.  In the Stirling approximation, $\log x! = x \log x + O(\log x)$, the log of the multinomial coefficient is

%
\begin{equation}\label{H:def}
   \log \mathbf{n!} = -\sum_i n_i \log \frac{n_i}{N} \doteq H(\mathbf{n}). 
\end{equation}
%

\noindent It is a concave functional  of $\mathbf{n}$ with functional derivatives 

%
\begin{equation}   \frac{\partial H(\mathbf{n})}{\partial n_i} = -\log\frac{n_i}{N} .
\end{equation}
%

\noindent It is also homogeneous in $\mathbf{n}$ with degree 1 and satisfies the Euler condition

%
\begin{equation}   H(\mathbf{n}) = \sum n_i \frac{\partial H(\mathbf{n})}{\partial n_i}. 
\end{equation}
%

\noindent Setting $p_i = n_i/N$ and applying $H$ to vector $\mathbf{p}$ we obtain
%
\begin{equation}   H(\mathbf{p}) = -\sum_i p_i \log p_i . 
\end{equation}
%

\noindent In this form $H$ reverts to the familiar entropy functional, historically associated with Boltzmann, Gibbs and Shannon. We will call it \textit{Shannon functional} and avoid the generic term ``entropy,'' whose meaning across disciplines varies. For our purposes the Shannon functional is defined as

%
\begin{equation}   H(\mathbf{a}) = H(a_1,a_2\cdots) = -\sum_i a_i \log \frac{a_i}{\sum _k a_k}
\end{equation}
%

\noindent and may be applied to any vector $\mathbf{a}$ with non-negative elements regardless of normalization.

\section{The Selection Functional}

Functional $W(\mathbf{n})$ biases the statistical weight of distribution $\mathbf{n}$ relative to its combinatorial multiplicity. We call it selection functional because it effectively selects distributions relative to each other. The functional derivative of $\log W$ is

%
\begin{equation}\label{wi:def}
   \log w_{i;\mathbf{n}} = \frac{\partial \log W(\mathbf{n})}{\partial n_i} ,
\end{equation}
%

\noindent and defines the cluster function $w_{i;\mathbf{n}}$, a property cluster mass $i$ in distribution $\mathbf{n}$. The cluster function $w_{i;\mathbf{n}}$ depends not only on $i$ but also on the distribution $\mathbf{n}$ on which this factor is evaluated. In the special case that $\log W$ is linear functional of $\mathbf{n}$ the functional derivative is a function of $i$ alone and is the same in all distributions. This special condition is associated with Gibbs distributions, which are discussed in Section \ref{sct:gibbs}. 

If $W(\mathbf{n})=1$ for all distributions, then the probability of distribution is proportional to its combinatorial multiplicity $\mathbf{n!}$. If this special condition is met we will call the ensemble \textit{unbiased}. The partition function of the unbiased ensemble can be easily determined by a combinatorial argument: it is equal to number of ways to assign $M$ objects into $N$ groups and is given by \citep{Matsoukas:springer_2019}

%
\begin{equation}\label{Omega:unbiased}
   \Omega_{M,N}^\circ = \binom{M-1}{N-1} . 
\end{equation}
%

\noindent Accordingly, the probability of distribution in this special case is

%
\begin{equation}   P^\circ(\mathbf{n}) = \mathbf{n!}\left/\binom{M-1}{N-1}\right..
\end{equation}
%

\noindent In a population undergoing transformations, for example aggregation, fragmentation etc.,  the selection functional is determined by the kinetic details of the mechanisms that produce these transformations; in the case of aggregation it is determined by the aggregation kernel $K_{i,j}$. The question arises whether  the unbiased ensemble is a possible solution of the Smoluchowski ensemble under some kernel.  The answer is yes, and is given in Section \ref{sct:gibbs}.

\section{Propagation Equations}

At the initial state all clusters are monomers and the distribution is $n_{i,0} = M\delta_{i,1}$. We set $W(\mathbf{n}^0)=1$ and since $\mathbf{n}^0!=1$ we also have $\Omega_{M,M} = 1$. We insert Eq.\ (\ref{Prob:n}) into Eq.\ (\ref{master}) and express the summation over parents of $\mathbf{n}$ as a summation over all pairs $(i-j,j)$ that produce mass $i$ in distribution $\mathbf{n}$. The result is (see Supplementary Material)

%
\begin{equation}\label{recursion_Omega:W}
   \frac{\Omega_{M,N+1}}{\Omega_{M,N}} 
   = 
   \left(\frac{M-N}{N} \frac{1}{\ens{K}_{M,N+1}}\right)   
   \left(
   \sum_{i=2}^\infty
   \frac{n_i}{M-N}
   \sum_{j=1}^{i-1} \frac{K_{i-j,j}}{\ens{K}_{M,N+1}}
   \frac{W(\mathbf{n'})}{W(\mathbf{n})}
   \right) .
\end{equation}
%

\noindent Here $N$ is the number of clusters in distribution $\mathbf{n}$ of generation $g=M-N$, $\mathbf{n'}$ is the parent distribution via the transition $(i-j)+(j)\to (i)$ and $\ens{K_{M,N+1}}$ is the mean kernel in the parent generation  $g'=g-1$. 
The left-hand side of Eq.\ (\ref{recursion_Omega:W}) depends solely on $M$ and $N$ whereas the second term on the right-hand side contains functionals of distribution $\mathbf{n}$. This term must be the same for all distributions $\mathbf{n}$ in the same generation in order to produce a result that is a function of $M$ and $N$ alone. From Eq.\ (\ref{Prob:n}) it is clear that $W$ and $\Omega_{M,N}$ may be defined within a proportionality constant $\alpha_{M,N}$; as long as this constant is common for all distributions in a generation it has no effect on probabilities and may be chosen arbitrarily. We choose it to satisfy the following criterion: if $W=\text{constant}$ for all distributions, we require this constant to be 1.  The choice that satisfies this condition is to set the double summation in Eq.\ (\ref{recursion_Omega:W}) to 1.  Equation (\ref{recursion_Omega:W}) now splits into two separate recursions, one for the partition function,

%
\begin{equation}
\label{propagation:sol}
   \frac{\Omega_{M,N+1}}{\Omega_{M,N}} 
   = \frac{M-N}{N} \frac{1}{\ens{K}_{M,N+1}}  
\end{equation}
%

\noindent and one for the selection functional,

%
\begin{equation}\label{recursion:W1}
   \sum_{i=2}^\infty
   \frac{n_i}{M-N}
   \sum_{j=1}^{i-1} 
   \frac{W(\mathbf{n'})}{W(\mathbf{n})}K_{i-j,j} = 1 . 
\end{equation}
%

\noindent The recursion for the partition function is readily solved to produce the partition function in generation $g=M-N$:

\begin{empheq}[box=\mybluebox]{equation}
\label{Omega:agg}
    \Omega_{M,N} = \Omega_{M,N}^\circ\prod_{\gamma=0}^{M-N+1}\ens{K_{M,M-\gamma}} .
\end{empheq}

\noindent Accordingly, the partition function is equal to the unbiased partition function times the product of all mean kernels from generation 0 up to the parent generation $g-1$. 
We write the recursion for the selection functional in the form

\begin{empheq}[box=\mybluebox]{equation}
\label{recursion:W}
   W(\mathbf{n}) = 
   \sum_{i=2}^\infty
   \frac{n_i}{M-N}
   \sum_{j=1}^{i-1} K_{i-j,j}
   W(\mathbf{n'}) . 
\end{empheq}
The result gives the selection functional of the offspring as a linear combination of selection functionals of all its parents. In principle this can be solved recursively for any distribution in any generation. For certain special cases the recursion can be solved in closed form. These are discussed in Section \ref{sct:gibbs}.

\chapter{Scaling Limit}\label{sct:thl}
\section{Most Probable Distribution}

We define the scaling limit by the condition $M,N\to\infty$ at fixed $M/N=\xav$. The expectation is that in this limit the intensive mean distribution $\ens{n_k}/N$ must converge to a limiting distribution $\bar p_k$ that is independent of $M$ and $N$ and depends only on $M/N=\xav$:

%
\begin{equation}\label{intensive:mpd}
   \frac{\ens{n_k}}{N} \to \bar p_k . 
\end{equation}
%

\noindent We further anticipate that the probability of distribution $P(\mathbf{n})$ becomes infinitely sharp around a single distribution, $\mathbf{n^*} = N \mathbf{p^*}$, such that $p^*_k$ is not merely the most probable distribution, it is overwhelmingly more probable than any other distribution in the ensemble. This further implies that the mean distribution and most probable distribution converge to each other:

%
\begin{equation}   \ens{p_k} \to p_k^* .
\end{equation}
%

\noindent This convergence is an implicit requirement for the validity of the Smoluchowski equation: the mean-field approximation is exact if a single distribution is representative of the entire ensemble. This is possible only if $P(\mathbf{\mathbf{n}})$ peaks very sharply about the most probable distribution. 
When a single term dominates the summation that defines the partition function in Eq.\ (\ref{Omega:def}), the log of the sum converges to the log of the maximum term,
\begin{empheq}[box=\mybluebox]{equation}
\label{logOmega:H:logW}
\vstrutSS
    \log \Omega_{M,N} 
    = H(\mathbf{n^*}) + \log W(\mathbf{n^*}) , 
\end{empheq}

\noindent with $H(\mathbf{n^*}) = \log \mathbf{n^*!}$.  As a further consequence of the intensive convergence in (\ref{intensive:mpd})  we have the Euler relationship for $\log W$:

\begin{empheq}[box=\mybluebox]{equation}
\vstrutSS
\label{logW:homo}
   \log W(\mathbf{n^*})
   =
   \sum_i n^*_i \log w^*_i .
\end{empheq}

\noindent where $\log w_i^* = \log w_{i;\mathbf{n^*}}$ is the functional derivative of $\log W(\mathbf{n^*})$,

%
\begin{equation}  \log w^*_i 
  = \frac{\partial\log W(\mathbf{n^*})}{\partial n^*_i} .
\end{equation}
%

\noindent Equation (\ref{logW:homo}) expresses the fact that $\log W$ is homogeneous functional of the \mpd. This condition follows from Eq.\ (\ref{logOmega:H:logW}) and the homogeneity properties of $H(\mathbf{n^*})$ and $\log\Omega_{M,N}$.

The most probable distribution (\mpd) maximizes the probability in Eq.\ (\ref{Prob:n}) among all distributions that satisfy the constraints in Eq.\ (\ref{constraints}). By Lagrange maximization we obtain the \mpd\ in the form

\begin{empheq}[box=\mybluebox]{equation}
\label{mpd}
    \frac{n^*_k}{N}
    =
    w^*_k \frac{e^{-\beta i}}{q} , 
\end{empheq}

\noindent and  $q$ and $\beta$ are parameters related to the Lagrange multipliers. We insert the \mpd\ into Eq.\ (\ref{logOmega:H:logW}) to obtain 

\begin{empheq}[box=\mybluebox]{equation}
\vstrutSS
\label{fundamental}
    \log\Omega_{M,N} = \beta M + (\log q) N . 
\end{empheq}

\noindent This fundamental equation relates the partition function to the primary variables of the ensemble: the macroscopic variables $(M,N)$ that define the ensemble,  and the Lagrange multipliers $(\beta,q)$ that appear in the \mpd. 
The convergence of $n^*_k/N$ to intensive limit $p^*_k$ implies that $\beta$ and $q$ are intensive, i.e., they are functions of $\xav=M/N$ but not of $M$ or $N$ individually. This further implies that Eq.\ (\ref{fundamental}) is homogeneous function of $M$ and $N$ with degree 1 and thus must satisfy Euler's theorem:

%
\begin{equation}   \log\Omega_{M,N} = 
   \left(\frac{\partial\log\Omega_{M,N}}{\partial M}\right) M
  +\left(\frac{\partial\log\Omega_{M,N}}{\partial N}\right) N .
\end{equation}
%

\noindent Direct comparison with Eq.\ (\ref{fundamental}) leads to:

\begin{empheq}[box=\mybluebox]{gather}
\label{beta}
    \beta  = \left(\frac{\partial\log\Omega_{M,N}}{\partial M}\right)_N,\\
\label{logq}    
    \log q = \left(\frac{\partial\log\Omega_{M,N}}{\partial N}\right)_M .
\end{empheq}

\noindent Thus the Lagrange multipliers that appear in the \mpd\ are the partial derivatives of the partition function. 
Differentiation of Eq.\ (\ref{fundamental}) with respect to all variables that appear on the right-hand side gives

%
\begin{equation}\label{Gibbs:Duhem}
   M d\beta + N d\log q = 0 .
\end{equation}
%

\noindent This is the Gibbs-Duhem equation associated with the Euler equation for $\log\Omega_{M,N}$ in Eq.\ (\ref{fundamental}). It may be written as

\begin{empheq}[box=\mybluebox]{equation}
\label{xav:logq:beta}
    \xav = - \frac{d\log q}{d\beta}   . 
\end{empheq}

\noindent In this form its expresses the relationship between $\beta$, $q$ and $\xav$.
%

The \mpd\ maximizes the log of the microcanonical weight, $H(\mathbf{n}) + \log W(\mathbf{n})$ and its maximum is $\log\Omega_{M,N}$. Therefore we have the inequality:

%
\begin{empheq}[box=\mybluebox]{equation}
\label{2nd:law:extensive}
\vstrutSS
    \log \Omega_{M,N} 
    \geq H(\mathbf{n}) + \log W(\mathbf{n}) .
\end{empheq}
%

\noindent It is satisfied by all distributions $\mathbf{n}$ in the $(M,N)$ ensemble with the equal sign only for $\mathbf{n}=\mathbf{n^*}$. This is the fundamental variational principle of the ensemble: it defines the \mpd\ and generates all relationships of this section. 

\section{Thermodynamics}
 
We recognize the equations of the previous section as those of familiar statistical thermodynamics.  Equation (\ref{mpd}) is the generalized canonical distribution, a member of the exponential family, whose parameters  $\beta$ and $q$ are related to the microcanonical partition function via Eqs.\ (\ref{fundamental}), (\ref{beta}) and (\ref{logq}). 
We define the extensive form of the canonical partition function $Q(\beta,N)$ via the Legendre transformation of $\log \Omega$:
\begin{equation}
   \log Q 
   = \log\Omega - M\left(\frac{\partial\log\Omega }{\partial M}\right)_N
   = N \log q,
\end{equation}
and thus we recognize $q = Q^{1/N}$ as the intensive form of the canonical partition function. 

The variational condition that produces the set of thermodynamic relationships is the inequality in Eq.\ (\ref{2nd:law:extensive}), which defines the \mpd\ as the distribution that maximizes the microcanonical weight. Expressing $H(\mathbf{n^*})$ and $\log W(\mathbf{n^*})$ in terms of the Euler relationships (\ref{H:def}) and (\ref{logW:homo}), respectively, this inequality takes the form

\begin{empheq}[box=\mybluebox]{equation}
\label{2nd:law}
   \frac{\log\Omega_{M,N}}{N} 
   \geq - \sum_i  p_i \log \frac{p_i}{w^*_i} ,
\end{empheq}

\noindent where $p_i=n_i/N$. The inequality is satisfied by all distributions $p_i$ with mean $\xav = M/N$ and the equality applies only to $p_i=p^*_i$. With $w_i^*=1$ it reduces the second law: the log of the microcanonical partition function is equal to the Shannon entropy of the most probable distribution, and this is larger than the entropy of any other distribution with the same mean. 

Table \ref{tbl:thermo} summarizes these relationships. They are consequences of the maximization of the probability in Eq. (\ref{Prob:n}) and are independent of the details of aggregation. These details enter only through Eqs.\ (\ref{Omega:agg}) and (\ref{recursion:W}), which express the partition function and the selection functional in terms of the aggregation kernel.

\begin{table} 
\newcommand{\myStrut}{\rule[-15pt]{0pt}{15pt}}
\caption{Summary of thermodynamic relationships}
\label{tbl:thermo}
\begin{equation*}
\renewcommand{\arraystretch}{3}
\begin{array}{c| *2{>{\displaystyle}l}p{5cm}}
\toprule
      \text{Most Probable Distribution} 
    & \frac{n^*_k}{N} = w^*_k \frac{e^{-\beta k}}{q} 
    & \text{Eq.\ (\ref{mpd})}
      \myStrut
\\\toprule
      \multirow{3}{*}{Partition Function}
    & \Omega_{M,N} = \beta M + (\log q) N  
    & \text{Eq.\ (\ref{fundamental})}
    \\
    & \beta=\left(\frac{\partial\log\Omega}{\partial M}\right)_N
    & \text{Eq.\ (\ref{beta})} 
    \\
    & \log q=\left(\frac{\partial\log\Omega}{\partial M}\right)_M
    & \text{Eq.\ (\ref{logq})}
      \myStrut
\\\toprule
   \text{Gibbs-Duhem Equation}
   & M d\beta + N d\log q = 0  
   & \text{Eq.\ (\ref{Gibbs:Duhem})} 
      \myStrut
\\\toprule
   \parbox[c][50pt][c]{100pt}
   {\centering Variational Condition\\ (Second Law)} & 
    \frac{\log\Omega_{M,N}}{N} 
    \geq -\sum_i p_i \log \frac{p_i}{w^*_i}  &
   \text{Eq.\ (\ref{2nd:law})}   
\\\toprule
\end{array}
\end{equation*}
\end{table}

\chapter{Gibbs Distributions}\label{sct:gibbs}
A special type of functional is of the form 

%
\begin{equation}\label{W:lin}
   W(\mathbf{n}) = \prod_i w_i^{n_i} ,
\end{equation}
%

\noindent whose log is linear in $\mathbf{n}$ 

%
\begin{equation}   \log W(\mathbf{n}) = \sum_i n_i \log w_i 
\end{equation}
%

\noindent with functional derivative $\log w_i$. Here $w_i$ is a function of $i$ alone and does not depend on $\mathbf{n}$. 
If the selection functional is given by Eq.\ (\ref{W:lin}) the probability of distribution in Eq.\ (\ref{Prob:n}) takes the form

\begin{empheq}[box=\mybluebox]{equation}
    P(\mathbf{n}) = \frac{N!}{\Omega_{M,N}} \prod_i \frac{w_i^{n_i}}{n_i!} .
\end{empheq}

\noindent 
Probability distributions of this type are called Gibbs distributions \citep{Berestycki:07} and are frequently encountered in stochastic processes \citep{Kelly:2011}. Several important results can be obtained in analytic form. In particular, the mean distribution is \citep{Matsoukas:springer_2019}:

\begin{empheq}[box=\mybluebox]{equation}
\label{gibbs:mean}
    \frac{\ens{n_k}}{N} = w_k \frac{\Omega_{M-k,N-1}}{\Omega_{M,N}} .
\end{empheq}

\noindent The result is exact for all $1\leq N\leq M$, $1\leq k \leq M-N+1$.  

We apply this selection functional of Eq.\ (\ref{W:lin}) to the transition $(i-j)+(j)\to (i)$ that converts parent distribution $\mathbf{n'}$ to offspring $\mathbf{n}$. By the stoichiometry of the transition we have

%
\begin{equation} 
  \frac{W(\mathbf{n'})}{W(\mathbf{n})} = \frac{w_{i-j} w_j}{w_i} .
\end{equation}
%

\noindent Inserting into Eq.\  (\ref{recursion:W1}) we obtain

%
\begin{equation}\label{recursion:W2}
   \sum_{i=2}^\infty
   \frac{n_i}{M-N}
   \sum_{j=1}^{i-1} 
   \frac{w_{i-j} w_j}{w_i} = 1 . 
\end{equation}
%

\noindent One possible solution that satisfies this equation for all distributions $\mathbf{n}$ is

%
\begin{equation}\label{wi:gibbs}
   w_i = \frac{1}{i-1}\sum_{j=1}^{i-1} w_{i-j} w_{j} K_{i-j,j};\quad
   w_1 = 1 , 
\end{equation}
%

\noindent This is not the only possible solution for $W$ in Eq.\ (\ref{recursion:W1}) and may or may not be acceptable; if it is, we  have obtained a Gibbs distribution and the kernel is a Gibbs kernel. 

We have identified three kernels for which Eq.\ (\ref{wi:gibbs}) is the correct solution. These are the constant kernel,

%
\begin{equation}   K_{i,j} = 1 ,
\end{equation}
%

\noindent the sum kernel

%
\begin{equation}   K_{i,j} = \frac{i+j}{2} 
\end{equation}
%

\noindent and their linear combinations. The product kernel is a \textit{quasi-Gibbs} kernel and is discussed in Section \ref{sct:prod}. 

Here we provide detailed solutions for the constant and sum kernels. We will not discuss the linear combination in part because the results are more involved but mainly because this kernel reverts to the sum kernel when cluster masses are large thus it does not contribute to our understanding of aggregation beyond what we learn by studying the constant and sum kernels separately. 

\section{Constant Kernel}

With $K_{i,j}=1$ Eq.\ (\ref{wi:gibbs}) gives $w_i=1$ for all $i$.  Accordingly, the  ensemble is \textit{unbiased} and its partition function is given by Eq.\ (\ref{Omega:unbiased}):

%
\begin{equation}\label{const:Omega}
   \Omega_{M,N} = \Omega_{M,N}^\circ = \binom{M-1}{N-1} . 
\end{equation}
%

\noindent The mean distribution follows from Eq.\ (\ref{gibbs:mean}) and is given by

%
\begin{equation}
\label{constK:mean:dstr}   
   \frac{\ens{n_k}}{N} = \left.
   \binom{M-k-1}{N-2}\right/
   \binom{M-1}{N-1} .
\end{equation}
%

\noindent To obtain the most probable distribution we calculate the parameters $\beta$ and $q$ from Eqs.\ (\ref{beta}) and (\ref{logq}) along with (\ref{const:Omega}). The differentiations may be done by first replacing the factorials in the partition function with the Stirling expression. Alternatively we may obtain these parameters by the discrete difference form of these derivatives and apply the asymptotic conditions $M,N\gg 1$. The latter method is simpler:

\begin{gather}
   \beta  
   = \log \frac{\Omega_{M+1,N}}{\Omega_{M,N}} 
   = \frac{M}{M-N+1} \to \frac{\xav}{\xav-1} ,  
   \\
   q = \frac{\Omega_{M,N+1}}{\Omega_{M,N}} = \frac{M-N}{N} =\xav - 1 .
\end{gather}

\noindent We obtain the \mpd\ from (\ref{mpd}) with $w^*_k=w_k$:

\begin{empheq}[box=\mybluebox]{equation}
\label{const:mean}
   \frac{n^*_k}{k} = \frac{1}{\xav-1}\left(\frac{\xav}{\xav-1}\right)^{-k} .
\end{empheq}

\noindent For large $\xav$ this goes over to the exponential distribution

%
\begin{equation}\label{const:mpd:cont}
   f(x) = \frac{e^{-x/\xav}}{\xav} , 
\end{equation}
%

\noindent which is the well known result for the constant kernel. Here $x$ stands for the continuous cluster mass.

\section{Sum Kernel}
The ensemble average of the sum kernel is

%
\begin{equation}   \ens{K}_{M,N} = \frac{M}{N} .
\end{equation}
%

\noindent We obtain the partition function from Eq.\ (\ref{Omega:agg}). The result is

%
\begin{equation}\label{sum:Omega}
   \Omega_{M,N} = N! \frac{M^{M-N}}{M!}\binom{M-1}{N-1} .
\end{equation}
%

\noindent The factors $w_i$ that satisfy Eq.\ (\ref{wi:gibbs}) are

%
\begin{equation}\label{sum:wi}
   w_k = \frac{k^{k-1}}{k!}  
\end{equation}
%

\noindent and the mean distribution follows from (\ref{gibbs:mean}),

%
\begin{equation}\label{sum:mean}
   \frac{\ens{n_k}}{N}
   =
   \frac{k^{k-1}}{k!}
   \frac{(M-k)^{M-N-k}}{M^{M-N-1}}
   \frac{(N-1)(M-N)!}{N(M-N-k+1)!} .
\end{equation}
%

\noindent This is an exact result for all $M\geq N\geq1$, $1\leq k \leq M-N+1$. The parameters $\beta$ and $q$ are obtained similarly to those for the constant kernel:

\begin{gather}
   \beta = 
     \frac{\Omega_{M+1,N}}{\Omega_{M,N}} \to 
     \frac{M-N}{M} - \log\frac{M-N}{M} , \\
   q = 
     \frac{\Omega_{M,N+1}}{\Omega_{M,N}} \to 
     \frac{M-N}{M}  .
\end{gather}

\noindent Combining with Eq.\ (\ref{mpd}) we obtain the \mpd\ in the form

\begin{empheq}[box=\mybluebox]{equation}
\label{sum:mpd}
    \frac{n_k^*}{N} = 
    \frac{k^{k-1}}{k!} \theta^{k-1} e^{k\theta} ,
\end{empheq}

\noindent with $\theta = 1-1/\xav$. We use the Stirling formula for the factorial the \mpd\ in the continuous limit takes the form

%
\begin{equation}\label{sum:mpd:cont} 
   f(x) = 
   \frac{\theta^{x-1}} {\sqrt{2\pi}}
   \frac{e^{-x\theta}} {x^{3/2}}.
\end{equation}
%

\begin{figure}
\begin{center}
\includegraphics[width=3.25in]{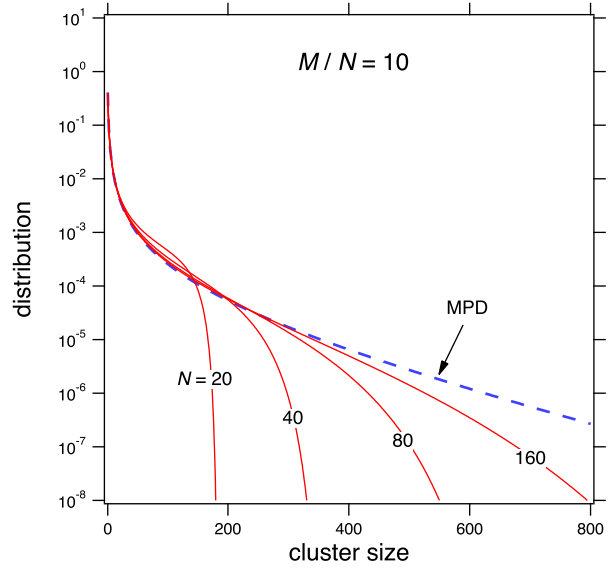}
\end{center}
\caption{Approach to scaling limit for the sum kernel at $\xav = 10$ $(\theta=0.9)$. The \mpd\ is calculated from Eq.\ (\ref{sum:mpd}) and the mean distribution from Eq.\ (\ref{sum:mean}) with $M=\xav N$, $N=20,40,80,160$. }
\label{fig2}
\end{figure}

\noindent Figure (\ref{fig2}) shows the \mpd\ for $\xav = 10$ and the mean distribution from Eq.\ (\ref{sum:mean}) at fixed $M/N=10$ for various values of $M$ and $N$. In the scaling limit the mean distribution converges to the $\mpd$.

\begin{table}[t]
\caption{Summary of Constant, Sum and Product Kernel; in all cases $\theta=1-1/\bar x$. }
\label{tbl:summary}
\[
\renewcommand{\arraystretch}{3}
\begin{array}{ c| *3{>{\displaystyle}c}p{5cm}}
 \makebox[80pt][c]{}          
        &  \makebox[80pt][c]{Constant Kernel} 
        &  \makebox[80pt][c]{Sum Kernel} 
        &  \makebox[80pt][c]{Product Kernel$^\ddagger$} 
       \\
\bottomrule
K_{i,j} & 1
        & (i+j)/2
        & i j
        \\
\Omega  & \binom{M-1}{N-1}
        & N! \frac{M^{M-N}}{M!}\binom{M-1}{N-1}
        & \left(N! \frac{M^{M-N}}{M!}\right)^2\binom{M-1}{N-1} 
       \\
\beta   & -\log\theta
         & \theta-\log \theta
        & 2\theta-\log \theta
       \\
q       & \frac{\theta}{1-\theta}
        & \theta
        & \theta(1-\theta)
       \\
w_k     & 1
        & \frac{k^{k-1}}{k!} 
        & \frac{2^{k-1} k^{k-2}}{k!} 
       \\
\mpd\   & (1-\theta) \theta ^{k-1}
        & \frac{k^{k-1}}{k!} \theta^{k-1} e^{k\theta} 
        & \frac{(2\theta k)^{k-2}}{k!}\frac{2\theta}{1-\theta}e^{-2\theta k}
\\
\bottomrule
\multicolumn{4}{l}{\text{$^\ddagger$Valid only for $\theta\leq 1/2$.} }
\end{array}
\]
\end{table}
\section{Quasi-Gibbs Kernels -- The Product Kernel}\label{sctn:prod}
\label{sct:prod}
We are able to obtain closed-form expressions for the partition function of the constant and sum kernels and heir linear combinations because they all satisfy the condition

%
\begin{equation}\label{gibbs_condition}
   \ens{K}_{M,N} = K(\mathbf{n}) 
\end{equation}
%

\noindent for all $\mathbf{n}$. This states that the mean kernel is the same in all distributions of the ensemble, therefore also equal to the ensemble average kernel. In this case the calculation of the ensemble average kernel is trivial and does not require knowledge of the probabilities $\mathbf{n}$. The constant kernel, sum kernel and their linear combinations are the only kernels that satisfy (\ref{gibbs_condition}) in the strictest sense, i.e., for all $\mathbf{n}$ that satisfy the two constraints in (\ref{constraints}). We refer to Eq.\ (\ref{gibbs_condition}) as the Gibbs condition because it generates Gibbs distributions. 
We may relax the requirement that \textit{all} distributions obey the Gibbs condition with the milder requirement that it be obeyed by \textit{most} distributions.  This is the case of the produce kernel. The product kernel is defined

\begin{empheq}[box=\mybluebox]{equation}
    K_{i,j} = i  j ,
\end{empheq}

\noindent and its mean within distribution $\mathbf{n}$ is

%
\begin{equation}   K(\mathbf{n}) = 
   \frac{N}{N-1}\left(\ens{i}^2 - \frac{\ens{i^2}}{N}\right) .
\end{equation}
%

\noindent Here $\ens{i} = M/N$ and $\ens{i^2}$ are the normalized first moment and second moments of $\mathbf{n}$, respectively. In the limit $N\to\infty$, $M/N=\text{fixed}$, this scales as

%
\begin{equation}\label{prod:Kn}
   K(\mathbf{n}) \sim \ens{i}^2 = \left(\frac{M}{N}\right)^2 ,
\end{equation}
%

\noindent in most distributions except those that contain clusters of the order $M$.\footnote{The largest cluster size in the ensemble is $\kmax=M-N+1$ and for $M\gg N$ it is of the order $M$.}  
According to Eq.\ (\ref{prod:Kn}) the product kernel is a quasi-Gibbs kernel: it satisfies the Gibbs condition in Eq.\ (\ref{gibbs_condition}) asymptotically in most but not all feasible distributions. We proceed to obtain the Gibbs distribution of the product kernel and test its validity. 

Inserting (\ref{prod:Kn}) into (\ref{Omega:agg}) we obtain the partition function:

%
\begin{equation}\label{prod:Omega}
   \Omega_{M,N} = \left(N! \frac{M^{M-N}}{M!}\right)^2\binom{M-1}{N-1} .
\end{equation}
%

\noindent We complete the solution by  evaluating $w_i$ from Eq.\ (\ref{wi:gibbs}),

%
\begin{equation}   w_k = \frac{2^{k-1}k^{k-2}}{k!} .
\end{equation}
%

\noindent The mean distribution is obtained by inserting these results into Eq.\ (\ref{gibbs:mean}):

%
\begin{equation}\label{prod:mean}
   \frac{\ens{n_k}}{N}
   =
   \frac{2^{k-1}k^{k-2}}{k!}
   \frac{
   (M-k)^{2(M-N-k+1)}
   M!
   (M-N)!
   }{
   N
   M^{2(M-N)}
   (M-1)!
   (M-N-k+1)!
   }
   .
\end{equation}

%
\begin{figure}
\begin{center}
\includegraphics[width=\textwidth]{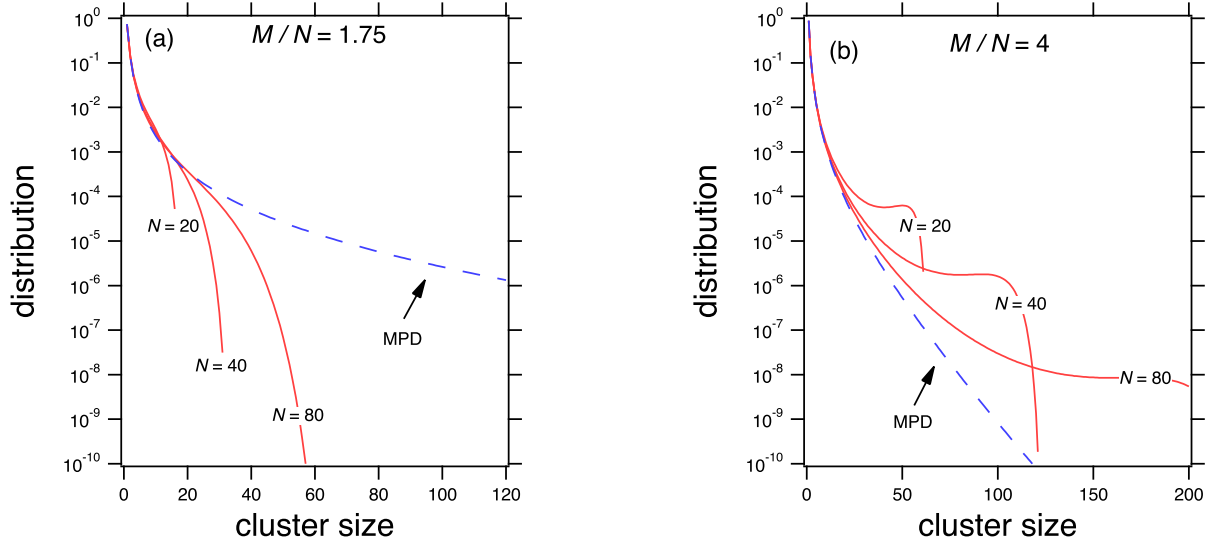}
\end{center}
\caption{Approach to the scaling limit for the product kernel with (a) $\xav = 1.75$ and (b) $\xav=4$. The MPD is calculated from Eq.\ (\ref{sum:mpd:cont}) and the mean distribution (dashed lines) from Eq.\ (\ref{prod:mean}).  The distributions for $\xav=4$ are not stable. }
\label{fig3}
\end{figure}

\noindent Unlike Eq.\ (\ref{const:mean}) and (\ref{sum:mean}) this result is not exact. This can be demonstrated  numerically by the fact this distribution is not normalized to unity and its mean is not $M/N$  for finite $M$, $N$; its approaches proper normalization in the asymptotic limit.  This failure arises from the fact that Eq.\ (\ref{gibbs:mean}) requires a Gibbs probability distribution that strictly applies to all distributions of the $(M,N)$ ensemble. 

We obtain $\beta$ and $\log q$ from Eqs.\ (\ref{beta}) and (\ref{logq}):

\begin{gather}
    \beta= \frac{M-N}{M} - 2\log\frac{M-N}{M} \\
    q = \frac{N(M-N)}{M^2} .
\end{gather}

\noindent Using $\theta = 1-1/\xav$ the \mpd\ is

%
\begin{equation}\label{prod:mpd}
   \frac{n^*_k}{N}
   =
   2^{k-1}
   \frac{(\theta k)^{k-2}}{k!}
   \frac{\theta}{1-\theta}
   e^{-2\theta k}
\end{equation}
%

\noindent and in the continuous limit

%
\begin{equation}\label{prod:mpd:cont}
   f(x)
   =
 \frac
    {2^{x} e^{x(-2 \theta)} \theta ^{x-1}}
    {\sqrt{8\pi } (1-\theta ) x^{5/2}} .
 \end{equation}
%

\noindent These results are summarized in Table \ref{tbl:summary} along with those for the constant and sum kernels. 

\noindent The relationship between the mean and the most probable distribution of the product kernel is shown in Fig.\ \ref{fig3} for two values of the mean cluster, $\xav=1.75$ and $\xav=4$. At $\xav=1.75$ the mean distribution calculated from Eq.\ (\ref{prod:mean}) is not exact but its moments asymptotically approach the correct values as $M$ and $N$ are increased at fixed $M/N=\xav$. At $\xav=4$ the behavior is different. A peak develops at the long tail of the distribution. It is pushed to ever larger sizes but never vanishes. In this region the mean distribution from Eq.\ (\ref{prod:mean}) is not correct: its mean does not converge to $\xav$ when $M$ and $N$ are increased, but to a value smaller than $\xav$, i.e., mass conservation is not satisfied. This breakdown is manifestation of \textit{gelation}, the emergence of an infinite cluster that is not captured by the mean field theory. The precise nature of the gel phase is discussed in the next section.

\begin{figure}
\begin{center}
\includegraphics[width=\textwidth]{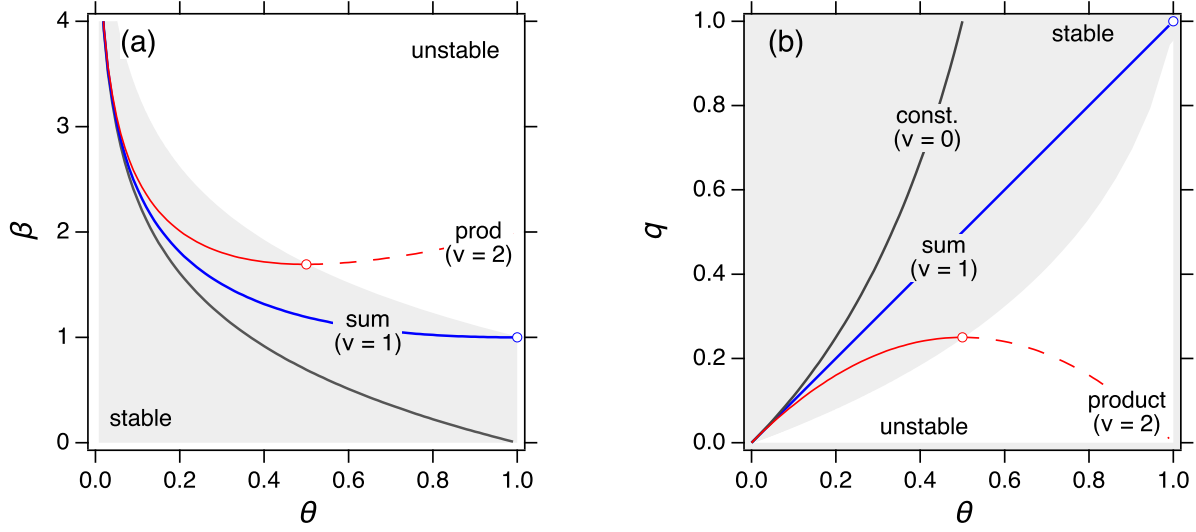}
\end{center}
\caption{Phase diagram of power-law kernels: In the shaded region  the system is stable and is represented by its \mpd. The unshaded region is unstable and the system is split into two phases, a sol phase and a gel phase, each represented by its own \mpd. Both graphs provide equivalent criteria of stability. }
\label{fig4}
\end{figure}

\chapter{Phase Behavior}\label{sct:stability}
\section{Stability}
The fundamental inequality of the ensemble is Eq.\ (\ref{2nd:law:extensive}) that defines the most probable distribution. This condition implies that the microcanonical functional is concave and this in turn implies that $\log\Omega_{M,N}$ is a concave \textit{function} of $M$ and $N$ and requires (see Supplementary Material)

%
\begin{equation}\label{stability:beta:logq}
   \frac{d\beta}{d\xav} \leq 0 \quad
   \text{or}\quad
   \frac{d\log q}{d\xav} \geq 0 . 
\end{equation}
%

\noindent These equivalent conditions guarantee the existence of the \mpd\ in the form of Eq.\ (\ref{mpd}). In thermodynamic language they ensure that the \mpd\ represents a stable state. 
The parameters $\beta$ and $q$ of the  constant, sum and product kernel are plotted in Figs.\ \ref{fig4}a and \ref{fig4}b, respectively, as a function of the progress variable $\theta = 1 - 1/\xav$. According to Eq.\ (\ref{stability:beta:logq}) stability requires $\beta$ to be decreasing function of $\xav$ and $q$ increasing function of $\xav$.  The constant kernel is stable at all $\theta$: $\beta_\text{const}$ decreases and $q_\text{const}$ increases monotonically over the entire range of $\theta$. 
The sum kernel is also stable at all $\theta$ but reaches the limit of stability at $\theta = 1$ or $\xav=\infty$. This kernel is borderline-stable: it is stable for all finite times and reaches instability at $t=\infty$. 
The product kernel is stable up to $\xav=0.5$ beyond which point both $\beta_\text{prod}$ and $q_\text{prod}$ violate the stability criteria. 

To survey the stability landscape of aggregation we employ the power-law kernel, 

%
\begin{equation}   K_{i,j} = (i j)^{\nu/2},
\end{equation}
%

\noindent with arbitrary exponent $\nu\geq 0$. This is a homogeneous kernel with degree $\nu$. It reverts to the product kernel with $\nu=2$ and to the constant kernel with $\nu=0$.  We treat this as a quasi-Gibbs kernel by analogy to the product kernel. We take the ensemble average power-law kernel to scale as

%
\begin{equation}\label{Kij_scaling}
   \ens{K}_{M,N} \sim \left(\frac{M}{N}\right)^\nu,
\end{equation}
%

\noindent and obtain the parameters $\beta$ and $q$ as

%
\begin{equation}\label{beta:q:powerlaw}
   \beta = \nu\theta -\log\theta,\quad
   q =\theta(1-\theta)^{\nu-1} . 
\end{equation}
%

\noindent With $\nu=0$ and $\nu=2$ these revert, as expected, to the results for the constant and product kernels, respectively. Interestingly, with $\nu=1$ we obtain the $(\beta,q)$ results for the sum kernel.  This  behavior turns the power-law kernel into a useful tool, a homogeneous kernel that reproduces the correct $(\beta,q)$ values of the constant, sum and product kernels, and which may be used to interpolate (and cautiously extrapolate) to other homogeneous kernels by varying the exponent $\nu$. 

The stability limit in power-law aggregation is reached at 

%
\begin{equation}   \theta^* = 1/\nu.
\end{equation}
%

\noindent Accordingly, the \mpd\ is stable in $0\leq\theta\leq \theta^*$ and unstable in $\theta^*<\theta\leq 1$. The phase diagram is shown in Figs.\ (\ref{fig4})a and (\ref{fig4})b with the stable region indicated by the shaded area. For $\nu\leq 1$ the system is stable at all $\theta$ from 0 to 1. For $\nu=1$ the limit of stability appears at $\theta=1$, which is reached in infinite time. In practice the system is stable at all finite times. For $\nu>1$ the  stability limit is reached within finite time at the point where the mean size reaches the critical value

%
\begin{equation}   \xav^* = \frac{1}{1-\theta^*} = \frac{\nu}{\nu-1} . 
\end{equation}
%

\noindent For $\nu=2$ (product kernel) the limit of stability is reached at $\xav^*=2$. We see from Fig.\ \ref{fig4} that both $\beta$ and $q$ reach the limit of stability simultaneously. 

\section{Phase Splitting -- The Sol-Gel Transition}
When the system crosses into the unstable region its state is no longer represented by the \mpd\ but by a mixture of two phases, each with its own \mpd. What are these phases?
To answer this question we begin with the observation that the elements of the ensemble are fundamentally \textit{discrete} distributions; the apparent continuity in the scaling limit is a mathematical artifact, a great convenience, but not a fundamental quality of the ensemble.  To understand the nature of the gel phase we must begin with a finite system. 
Given a distribution of $M$ particles partitioned into $N$ clusters, the maximum cluster mass possible is $\kmax = M-N+1$ and is found in a single distribution of the ensemble, in which one cluster contains $M-N+1$ units and the remaining $N-1$ clusters contain one unit mass each. 
The region $(\kmax+1)/2 < k \leq \kmax$ is special: it is either empty, or it contains a single cluster. It cannot contain more than one cluster because there is not enough mass to have two clusters that are both larger than $(\kmax+1)/2$. In the event that it does contain a cluster, its mass is of the order of $\kmax = M-N+1$, and in the asymptotic limit, of the order $M$. 
This means that the mass in the region $k>\kmax+1)/2$ is of the same order as that in $k<(\kmax+1)/2$. A cluster in $k>(\kmax+1)/2$ represents a \textit{giant component}, a single element of the population that carries a finite fraction of the total mass contained in the distribution. 

The set of distributions that do not contain a giant cluster constitute the \textit{sol} phase; sol distributions satisfy the scaling form of the mean kernel in Eq.\ (\ref{Kij_scaling}) and the Gibbs condition in Eq.\ (\ref{gibbs_condition}). 
Distributions that contain a cluster in the gel region violate the Gibbs condition and will be treated as a mixture of a sol phase $(k\leq (\kmax+1)/2)$ and a gel phase ($k>(\kmax+1)/2$). 
Given an individual distribution $\mathbf{n}$, a certain fraction of mass is contained in the sol region with the rest in the gel region. 
The ensemble averages of these fractions define, respectively, the sol fraction, $\phi_\sol$, and gel fraction, $\phi_\gel$, in the ensemble:

%
\begin{equation}   \phi_\sol =
   \frac{1}{M} 
   \sum_\mathbf{n} P(\mathbf{n}) 
   \sum_{k=1}^{k'} n_k ;
\quad
   \phi_\gel = 
   \frac{1}{M} 
   \sum_\mathbf{n} P(\mathbf{n})
   \sum_{k=k'+1}^{\kmax} n_k ;
\quad
   \phi_\sol+\phi_\gel = 1 , 
\end{equation}
%

\noindent with $k' = (\kmax+1)/2$. If $P(\mathbf{n})$ is such that in the scaling limit $\phi_\gel\to 0$, the ensemble consists of a single phase, the \textit{sol}, and is represented by the \mpd\ in Eq.\ (\ref{mpd}). If $\phi_\gel>0$ the ensemble is represented by a mixture of the two phases.  We will determine their distributions and construct the tie line between the two phases.  

We suppose that the state at $(M,N)$ consists of a sol phase with $M_\sol$, $N_\sol=N-1$, and a gel phase with $M_\gel=M-M_\sol$. The evolution of the sol phase is governed by Eq.\ (\ref{propagation:sol}), which we now write as

%
\begin{equation}\label{propagation:sol:powerlaw}
   \left(\frac{\Omega_{M+1,N}}{\Omega_{M,N}}\right)_\sol 
   = q(\theta_\sol) 
   = \theta_\sol(1-\theta_\sol)^{\nu-1} .
\end{equation}
%

\noindent This must be satisfied by the sol phase at all times.  
In the pre-gel region the state is a single phase, sol,  with $\theta_\sol=\theta = 1-N/M$ and  $q$-$\beta$ parameters from Eq.\ (\ref{beta:q:powerlaw}). 
In the post-gel region it is a mixture of two phases: a sol phase with mass $M_\sol$ and number of clusters $N_\sol=N-1$; and gel phase with mass  $M_\gel = M-M_\sol$ found in a single cluster ($N_\gel=N-1$). 
The sol phase is determined from Eq.\ (\ref{propagation:sol:powerlaw}) with $\theta_\sol = 1 - M_\sol/(N-1)$ and its $\beta$-$q$ parameters are given by Eq.\ (\ref{beta:q:powerlaw}) with $\theta=\theta_\sol$. The mass of the gel phase is then obtained from the conservation conditions $M_\gel = M-M_\sol$. 
These steps are summarized below.

\begin{itemize}
\item[] \textbf{Pre-Gel Region} $0\leq \theta < \theta^*$

The system consists of a sol phase and its \mpd\ is

%
\begin{equation}   \frac{n^*_k}{N} = w^*_k \frac{e^{-\beta(\theta)}}{q(\theta)} 
\end{equation}
%

\noindent with 

%
\begin{equation}   \beta = \nu\theta -\log\theta,\quad
   q =\theta(1-\theta)^{\nu-1},\quad
   \theta=1-N/M . 
\end{equation}
%

\noindent \item[] \textbf{Post-Gel Region} $\theta^*\leq \theta < 1$ The system consists of a sol phase with mass fraction $\phi_\sol$ and a gel phase with fraction $\phi_\gel=1-\phi_\sol$. 

\begin{enumerate}
\item Obtain $\theta_\sol$ by solving 

%
\begin{equation}\label{xsol:retrace}
   q(\theta_\sol) = q(\theta),\quad
   \theta_\sol\leq \theta^* .
\end{equation}
%

\noindent with $q(\theta)$ from Eq.\ (\ref{beta:q:powerlaw}) and $\theta = 1-1/\xav = 1-N/M$. 

\item Obtain $\phi_\sol$ and $\xav_\sol$ from

%
\begin{equation}   
   \phi_\sol = \frac{1-\theta}{1-\theta_\sol} , \quad
   \xav_\sol = \frac{1}{1-\theta_\sol} . 
\end{equation}
%

\item Obtain the gel fraction from mass balance:

%
\begin{equation}\label{phi:gel}
   \phi_\gel=1-\phi_\sol = \frac{\theta-\theta_\sol}{1-\theta_\sol} . 
\end{equation}
%

\noindent The mean size of the gel cluster is $\phi_\gel M$, where $M$ is the total mass in the system. In the scaling limit the gel fraction is 1 and the size of the gel cluster is $\infty$. 

\end{enumerate}
\end{itemize}

\begin{figure}
\begin{center}
\includegraphics[width=3.25in]{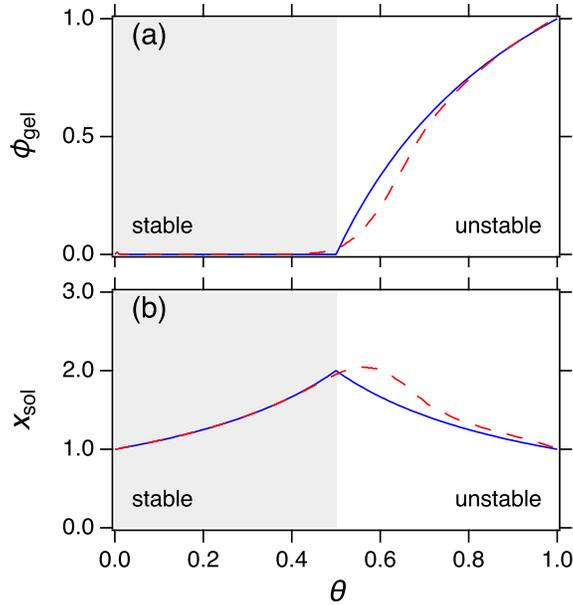}
\end{center}
\caption{(a) Gel fraction and (b) mean sol cluster size as a function of the progress variable $\theta$. Past the gel point the mean size in the sol retraces its pre-gel history back to its initial size $\xav_\sol=1$. The dashed lines are MC simulations with $M=200$ particles. }
\label{fig5}
\end{figure}

The gel fraction and the mean cluster size for the product kernel $(\nu=2)$ are shown in Fig.\ (\ref{fig5}) as a function of $\theta$. The gel fraction is zero up until the gel point ($\theta^*=0.5$) and increases according to Eq.\ (\ref{phi:gel}) once in the post-gel region. The mean cluster size increases in the pre-gel region but decreases in the post-gel region, as clusters in the sol are lost by reaction with the gel. At $\theta\to 1$ ($t\to\infty$) all mass is found in the gel phase except for a single sol particle with unit mass.  This is the infinite dilution limit of the sol phase, to borrow the terminology of solution thermodynamics. 

The evolution of $\xav_\sol$ past the gel point retraces its pre-gel history. This is a consequence of Eq.\ (\ref{xsol:retrace}), which resolves the sol phase in the two-phase region.  The symmetry of $q(\theta)$ about $\theta^*=0.5$ in the case of the product kernel produces a correspondingly symmetric evolution of $\xav_\sol$, as shown in Fig.\ \ref{fig5}b. The dashed lines are Monte Carlo simulations with $M=200$ particles and are shown for comparison (the simulations are discussed in the next section). The deviation from theory near the gel point is due to finite size effects. In these simulations a relatively small number of particles was used to permit the collection of a large number of realizations within reasonable computational time.

\begin{figure}
\begin{center}
\includegraphics[width=4.25in]{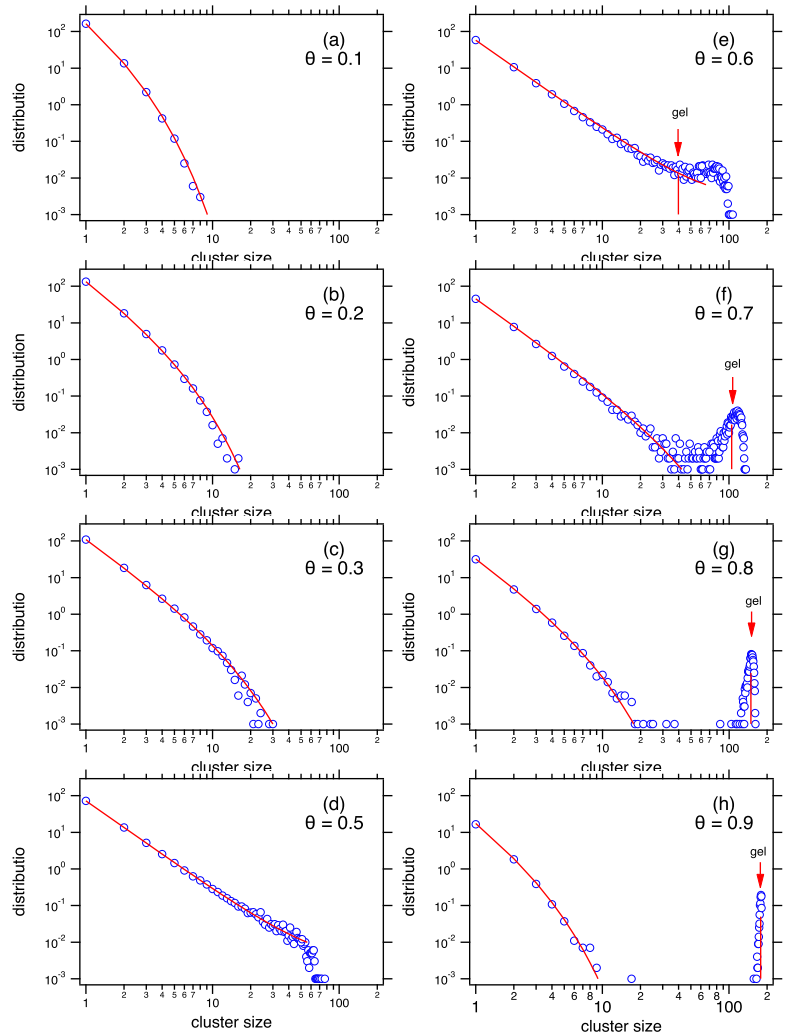}
\end{center}
\caption{Monte Carlo snapshots of the mean distribution of the product kernel with $M=200$ particles (open circles are MC results, solid lines are calculated from theory). The gel phase emerges at $\theta^*=0.5$ and moves towards ever larger sizes (arrows mark the theoretical predictions). The distribution of the sol grows in the pre-gel region range $0<\theta<0.5$ but contracts once past the post-gel point ($\theta>0.5$). }
\label{fig6}
\end{figure}

\section{Monte Carlo Simulations}
We place the theory into context via Monte Carlo (MC) simulation.  The simulation is implemented by the constant-$V$ Monte Carlo method \citep{Matsoukas:CES98} and tracks a sample of clusters that undergo binary aggregation with probability proportional to the transition rate $R_{i,j}$ in Eq.\ (\ref{transition_rate}). 
At each step the simulation box contains a sample of $N$ clusters, with $N$ decreasing from $M$ to 1 as clusters merge.  A pair of clusters are chosen at random and is combined into a single cluster according the following criterion: draw a random number $\text{rnd}$ in the interval $(0,1)$ and accept the merging of the clusters if

\begin{equation}
   \text{rnd} < \frac{K_{i,j}}{K_\text{max}}, 
\end{equation}

\noindent
where $K_{i,j}$ is the aggregation kernel between the chosen clusters and $K_\text{max}$ is the maximum aggregation kernel in the simulation box. If the criterion is satisfied the event is accepted and the reactant particles are deleted and replaced by a cluster with their combined mass. If the event is rejected, a new pair is chosen and the process is repeated. 
The simulation begins with $M$ monomers and continues until a single cluster is formed. This amounts to a random walk along the edges of the graph in Fig.\ \ref{fig1} that spans its entire range from $\theta=0$ to $\theta = 1-1/M$. A trajectory from the top to the bottom of the graph consists of a sequence of $M$ sampled distributions, one  from each generation. By averaging trajectories we obtain the mean distribution in each generation, which may then be compared to the mean distribution predicted by the theory. 

Figure (\ref{fig6}) shows the evolution of the mean distribution obtained by MC simulation  with the product kernel using $M=200$. Up until the gel point is reached the state is a single sol phase. It is characterized by a population of clusters whose tail decays fast enough  that its moments are finite. 
Above the gel point a gel peak emerges. It becomes more pronounced and moves to larger sizes as aggregation progresses. 
Past the gel point the sol distribution contracts and retraces its steps back to the monomeric state as $\theta$ increases. For example, the sol distribution at $\theta = 0.9$ is identical to that at $\theta=0.1$ except that it carries less mass. 
In the Smoluchowski literature this is known as the \textit{Flory} solution to gelation \citep{Ziff:JCP80}. A competing solution by Stockmayer \citep{Stockmayer:JCP43} predicts that the intensive distribution of the sol phase remains constant past the gel point except for the fact that its mass gradually decreases as it is transferred to the gel. As it turns out, Stockmayer solution implicitly assumes that $P(\mathbf{n})$ is strictly a Gibbs distribution. In this case the sol-gel tie line is obtained by  equating the temperatures of the two phases and the sol distribution is indeed found to be constant throughout the post gel region  \citep{Matsoukas:statPBE_PRE14}. An  analysis of the Stockmayer solution is beyond the scope of this work but a commentary is given in \citep{Matsoukas:springer_2019}.

\chapter{Continuous Limit}\label{sct:cont}
We define the continuous limit by the conditions

\begin{equation*}
   M\geq N \to \infty,\quad
   \xav \gg 1.  
\end{equation*}

\noindent Thus in addition to the scaling limit we require the mean cluster size to be much larger than the unit mass, such that the cluster mass may be treated as a continuous variable, which we denote as $x$. Equations (\ref{const:mpd:cont}), (\ref{sum:mpd:cont}) and (\ref{prod:mpd:cont}) refer to this limit. We present the corresponding expressions for the partition function and the selection functional. 

In the continuous domain all intensive properties of the ensemble are functions of the mean cluster size $\xav$. Thus we write $\beta=\beta(\xav)$, $q = q(\xav)$, $w^*_i = w(x;\xav)$, and express the partition function in intensive form $\log\omega(\xav) = (\log\Omega_{M,N})/N$. 
The \mpd\ is

\begin{empheq}[box=\mybluebox]{equation}
\label{cont:mpd}
   f(x) = w(x;\xav) \frac{e^{- x\beta(\xav)}}{q(\xav)}  
\end{empheq}

\noindent and satisfies the normalizations

%
\begin{equation}   
   \int_0^\infty f(x) dx = 1,\quad
   \int_0^\infty x f(x) dx = \xav . 
\end{equation}
%

\noindent The log of the cluster function $w(x;\xav)$ is the functional derivative of the selection functional at the \mpd:

%
\begin{equation}   \log w(x;\xav) = \frac{\delta\log W[f]}{\delta f}
\end{equation}
%

\noindent and the notation $w(x;\xav)$ indicates this function of $x$ will generally depend on $\xav$ as well since the functional derivative of non linear functionals depend on the function on which the derivative is evaluated. 
Since the microcanonical probability peaks sharply about the \mpd\ (we are assuming a stable single-phase state) all ensemble averages revert to averages over the continuous \mpd. The ensemble average kernel is then  equal to the mean kernel within the \mpd\ 

%
\begin{equation}   \ens{K}_{M,N} \to \bar K(\xav) = 
   \int\limits_0^\infty dx
   \int\limits_0^\infty dy
   K(x,y) f(x) f(y) . 
\end{equation}
%

\noindent The log of the intensive partition function, $\log\omega(\xav) = \log\Omega_{M,N}/N$,  satisfies

\begin{equation}
 \label{fundamental:omega}
  \log\omega = \beta \xav + \log q, 
\end{equation}

\noindent with 

\begin{equation}
\label{beta:omega}
   \beta = \frac{d\log\omega}{d\xav} .
\end{equation}

\noindent These are the intensive forms of Eqs.\ (\ref{fundamental}) and (\ref{beta}), respectively. The partition function of aggregation is  obtained from Eq.\ (\ref{Omega:agg}) by expressing the summation over $\log\ens{K}_g$ and an integral over $\bar K(\xav)$:

\begin{empheq}[box=\mybluebox]{equation}
\label{cont:log:omega}
    \log\omega 
    = 1+\log\xav
    + \xav \int_0^{\xav}\log\bar K(y) \frac{dy}{y^2} . 
\end{empheq}

\noindent The parameter $\beta$ is obtained from Eq.\ (\ref{beta:omega}) and $\log q$ from (\ref{fundamental:omega}):

\begin{empheq}[box=\mybluebox]{gather}
\label{cont:beta}
    \beta = \frac{1}{\xav} + \frac{\log \bar K(\xav)}{\xav}
          + \int_0^{\xav}\log\bar K(y) \frac{dy}{y^2}, \\
\label{cont:q}
     q = \frac{\xav}{\bar K(\xav)} . 
\end{empheq}

\noindent Finally we obtain the selection functional from Eq.\ (\ref{recursion:W}) setting $\mathbf{n}\to\mathbf{n^*}$ and converting the summations into integrals: 

%
\begin{equation}   \frac{W(\mathbf{n^*})}{W(\mathbf{n^{*'}})}
   = \frac{1}{M-N}\sum_{i=2}^\infty n^*_i\sum_{j=1}^{i-1} K_{i-j,j}
   \to 
   \frac{1}{x-1}\int_1^\infty dx f(x)
   \int_{1}^{x-1} dy K(x-y,y) . 
\end{equation}
%

\noindent where $\mathbf{n^*}$ is the \mpd\ at $\xav=M/N$ and $\mathbf{n^{*'}}$ is its parent at $\xav'=M/(N-1)$. In the limit $\xav-\xav'\to d\xav$ this becomes

\begin{empheq}[box=\mybluebox]{equation}
\label{cont:log:W}
    \frac{d\log W[f]}{d\xav}
    =
    \log\left(
    \frac{1}{x}
    \int_0^\infty dx f(x) \int_{0}^{x} dy K(x-y,y) 
    \right). 
\end{empheq}

\noindent Equations (\ref{cont:mpd}), (\ref{cont:log:omega}) (\ref{cont:beta}), (\ref{cont:q}) and (\ref{cont:log:W}) provide an equivalent mathematical description of Smoluchowski aggregation in the continuous limit. These are accompanied by the variational condition 
\begin{empheq}[box=\mybluebox]{equation}
\label{cont:2nd:law}
   \log\omega \geq -\int_0^\infty p(x)\log \frac{p(x)}{w(x;\xav)} dx,
\end{empheq}
which is the continuous form of Eq.\ (\ref{2nd:law}) and is satisfied by all distributions $p(x)$ with mean $\xav$. The equality defines the solution to the Smoluchowski process, the \mpd, $f(x)$. 

\textbf{Special Case -- Constant Kernel:} As a demonstration we apply these results to the constant kernel. The right-hand side of Eq.\ (\ref{cont:log:W}) is zero and we obtain $W[f]=1$ at all times. It follows that $w(x;\xav) = 1$.  The parameters $\beta$ and $q$ are

\begin{gather}
   \beta = 1/\xav,\quad
   q = \xav,
\end{gather}

\noindent and the \mpd\ becomes

%
\begin{equation}   f(x) = \frac{e^{-x/\xav}}{\xav} .
\end{equation}
%

\noindent This is the well-known solution of the constant kernel in the continuous domain. The partition function of the constant kernel is

%
\begin{equation}      \log\omega = 1+\log\xav ,
\end{equation}
%

\noindent and the inequality in Eq.\ (\ref{cont:2nd:law}) becomes

%
\begin{equation}   1 + \log\xav \geq -\int p(x)\log p(x) dx = H[g] ,
\end{equation}
%

\noindent whose right-hand side is the Shannon entropy. For fixed $\xav$, $x>0$ it is maximized by the exponential distribution whose entropy is $1+\log\xav$: the inequality is indeed satisfied.

\chapter{Summary}\label{sct:summary}
With the results obtained here we have made contact with several previous works in the literature. The mean distribution for the constant kernel in Eq.\ (\ref{constK:mean:dstr}) was given by \citet{Hendriks:ZPCM85A}, who also obtained a recursion for the partition function  similar to that in Eq.\ (\ref{propagation:sol}). 
The combinatorial treatment of Hendriks has in fact several common elements to ours. It is limited, however, to the constant and sum kernels and lacks the thermodynamic element of this work. The recursion for the cluster weights in Eq.\ (\ref{wi:gibbs}) has appeared in various treatments of aggregation, both deterministic \citep{Lushnikov:JCIS73,Leyvraz:PR03}
and stochastic  \citep{Spouge:1983,Spouge:M83_121,Hendriks:ZPCM85A}. The mean distributions in the continuous limit for the mean and sum kernels and for the product kernel in the pre-gel region are well known results in the literature \citep{Leyvraz:PR03}. The instability of power-law kernels has been discussed by \citet{Ziff:PRL82} based on the Smoluchowski equation.  These connections to prior literature serve to validate the theory presented here and demonstrate that the thermodynamic treatment provides a unified theory of aggregation that brings previously disconnected results under a single formalism,  the \textit{Smoluchowski ensemble}. 

The Smoluchowski ensemble is a probability space of distributions that are feasible under the rules of binary aggregation. The structure of this space, i.e., the connectivity of the graph  in Fig.\ (\ref{fig1}), is solely determined by the condition that aggregation is a binary event; the probability measure over this space is determined by the rate expression prescribed by the aggregation model.  In Smoluchowski aggregation the rate is directly proportional to the number of clusters that appear on the reactant side of the aggregation reaction and on the aggregation kernel.
In the scaling limit the probability of distribution is sharply peaked around a single distribution of the ensemble, its most probable distribution (\mpd). In this limit all ensemble averages reduce to averages of the \mpd\ and this distribution alone suffices to produce all properties of the ensemble. The Smoluchowski coagulation equation is the time evolution of the most probable distribution in the asymptotic limit. 

The step that turns the Smoluchowski ensemble into a \textit{thermodynamic} ensemble is Eq.\ (\ref{Prob:n}), which expresses the probability of distribution in the ensemble in terms of two special functionals, the multinomial coefficient $\mathbf{n!}$ and the selection functional $W(\mathbf{n})$. This formulation introduces the partition function $\Omega_{M,N}$ as the central property of the ensemble to which al other properties are connected. The thermodynamic calculus, summarized by the equations in Table \ref{tbl:thermo}, is a mathematical consequence of the variational condition that defines the most probable distribution in Eq.\ (\ref{mpd}) as the solution to the constrained maximization of the probability $P(\mathbf{n})$ in Eq.\ (\ref{Prob:n}). The constraints are given by Eqs.\ (\ref{constraints}) that fix the zeroth and first order moments of the distribution. These constraints define a \textit{microcanonical} ensemble of distributions with fixed mean $\xav = M/N$. 

The \mpd\ obtained by the method of constrained maximization is stable, provided that the partition function is concave in its independent variables. In extensive terms, $\log\Omega_{M,N}$ must be concave in $M,N$; in intensive terms, $\log\omega(\xav)$ must be concave in $\xav$. The two conditions are equivalent and define the stability criterion of the \mpd. As in regular thermodynamics, when the stability criterion is violated the system experiences \textit{phase splitting} and exists a mixture of two phases --mathematically, as a linear combination of two independent \mpd's. In aggregation these phases are the \textit{sol phase}, which is represented by the \mpd\ in Eq.\ (\ref{mpd}) and the \textit{gel phase} (giant component), which in the scaling limit is represented by a delta function at $\infty$. The splitting into a sol and gel phase is treated by the theory in a very natural and rigorous manner. 

Notably, entropy in this treatment plays no special role. The Shannon entropy of distribution is the log of the multinomial coefficient. In the scaling limit, entropy is a component of the partition function through equation (\ref{logOmega:H:logW}),

\begin{equation*}
   \log\Omega_{M,N} = H(\mathbf{n^*}) + \log W(\mathbf{n^*}), 
\end{equation*}

\noindent where $H(\mathbf{n^*})$ is the Shannon functional evaluated at the \mpd. In the special case of the constant kernel $W(n^*)=1$. In this case the partition function reduces to the Shannon entropy of the \mpd,

%
\begin{equation}   \log\Omega_{M,N} = -N \sum_i \frac{n_i}{N}\log \frac{n_i}{N};\quad 
   \text{(constant kernel)},
\end{equation}
%

\noindent and the variational condition reads,

%
\begin{equation}   H(\mathbf{n}) \leq H(\mathbf{n^*}) = \Omega_{M,N};\quad 
   \text{(constant kernel)} . 
\end{equation}
%

\noindent In this form we have recovered the inequality of the second law as stated in statistical thermodynamics: the entropy of the equilibrium distribution (\mpd) is at maximum with respect to all feasible distributions, namely, all distributions with the same mean. As is well known this distribution is exponential. The constant kernel is special. With $W(\mathbf{n})=1$ the probability of distribution is proportional to $\mathbf{n!}$; accordingly, all ordered sequences of $N$ clusters with total mass $M$ are equally probable. The ordered sequence of cluster masses in this case is analogous to microstate in statistical mechanics and the condition $W=1$ analogous to the postulate if equal a priori probabilities.  In the general case the Shannon entropy and the log of the microcanonical partition function are not the same. The fundamental functional that is maximized is the microcanonical weight $\mathbf{n!} W(\mathbf{n})$, whose log is 

\begin{equation*}
    H(\mathbf{n}) + \log W(\mathbf{n}) .
\end{equation*}

\noindent The selection functional incorporates the effect of the aggregation kernel and in this sense it the point of contact between thermodynamics and the mathematical model of the stochastic process that gives rise to the probability space of interest. In Smoluchowski aggregation the model is defined by the transition rate in Eq.\ (\ref{T:def}) and the corresponding governing equation for $W$ is Eq.\ (\ref{recursion:W}). 

The thermodynamic formalism developed here is not limited to aggregation. Two alternative derivations that make no reference to stochastic process that gives rise to the probability space have been given in \citep{Matsoukas:statPBE_PRE14} and \citep{Matsoukas:E19}. As long as $\log W$ is a homogeneous functional with degree 1, the thermodynamic relationships follow as a direct consequence of the maximization of the microcanonical probability in Eq.\ (\ref{Prob:n}) under the constraints in Eq.\ (\ref{constraints}). The details of aggregation enter through Eqs.\ Eq.\ (\ref{Omega:agg}) and (\ref{recursion:W}) that give the partition function and selection functional in terms of the aggregation kernel. The approach may be generalized to other processes including growth by monomer addition and breakup. These will be treated elsewhere.


\bibliographystyle{abbrvnat}
\bibliography{pbe,StatMech,tm}


\appendix
\chapter{Supplementary Material}
\setcounter{equation}{0}
\renewcommand{\theequation}{A\arabic{equation}}

\section{Derivation of the Smoluchowski equation}
Define 
\begin{equation}
\label{t:def}
   t_{i-j,j} =    \frac{K_{i-j,j}}{\ens{K}_{g-1}}
   \frac{n'_{i-j} (n'_j-\delta_{i-j,j})}{1+\delta_{i-j,j}} ,
\end{equation}
such that the transition probability rate is
\begin{equation}
   \frac{R_{i-j,j}}{\ens{R}_{g-1}} 
   = \frac{2 t_{i-j,j}}{N'(N'-1)}
   = \frac{2 t_{i-j,j}}{N(N+1)} ,
\end{equation}
where $N'$ and $N=N'-1$ is the number of clusters in the parent and offspring generation, respectively. Equation (\ref{master}) is now written as
\begin{equation}
\label{master:2}
   P(\mathbf{n}) = 
   \frac{2}{N(N+1)} \sum_{i=2}^{\infty}\sum_{j=1}^{i/2}
   P(\mathbf{n'}) t_{i-j,j} . 
\end{equation}
Each term of the summation represents an elementary aggregation event that produces an element of $\mathbf{n}$ from a parent via the transition $(i-j)+(j)\to (i)$. 
We substitute the stoichiometric relationship
\begin{equation}
   n'_i = n_k+\delta_{k,i} - \delta_{k,i-j} - \delta_{k,j} 
\end{equation}
into this result to obtain
\begin{equation}
\label{app:smoluchowski:1}
   \ens{n_k}_{M,N} = \frac{2}{N(N+1) \ens{K}_{M,N+1}}
   \ens{
   \sum_{i=2}^\infty
   \sum_{j=1}^{i-1}
   (n'_k+\delta_{k,i} - \delta_{k,i-j} - \delta_{k,j})
   t_{i-j,j}
   }
\end{equation}
and workout each term of the summation separately. The first term evaluates the mean number of clusters int he parent ensemble:
\begin{equation}
   \frac{2}{N(N+1)\ens{K}_{M,N+1}}
   \ens{
   n_k
   \sum_{i=2}^\infty
   \sum_{j=1}^{i-1}
   t_{i-j,j}
   }_{M,N+1}
   =\ens{n_k}_{M,N+1}
   .
\end{equation}
The second term is
\begin{equation}
   \frac{2}{N(N+1)\ens{K}_{M,N+1}}
   \ens{
   \sum_{j=1}^{k/2} t_{k-j,j}
   }_{M,N+1}
   =
   \frac{2}{N(N+1)\ens{K}_{M,N+1}}
   \ens{
   \sum_{j=1}^{k-1} n_{k-j}(n_j-\delta_{k-j,j}) K_{k-j,j}
   }
\end{equation}
and the third term is
\begin{equation}
   \frac{2}{N(N+1)\ens{K}_{M,N+1}}
   \ens{
   \sum_{j=1}^{\infty}(1+\delta_{k,j}) t_{k-j,j}
   }_{M,N+1}
   =
   \frac{2}{N(N+1)\ens{K}_{M,N+1}}
   \ens{
   n_k(n_j-\delta_{k,j})
   }
   .
\end{equation}
Inserting these results into (\ref{app:smoluchowski:1}) we obtain Eq.\ (\ref{smoluchowski_1}) in the main text.

\section{Derivation of Eq.\ (\ref{recursion_Omega:W})}
We begin with the propagation equation for $P(\mathbf{n})$ in Eq.\ (8) and substitute Eq.\ (11) to obtain
\begin{equation}
\label{app:master:1}
   P(\mathbf{n}) = 
   \sum_\mathbf{n'}
   P(\mathbf{n'})
   \frac{n'_{i-j}(n_j-\delta_{i-j,j})}{1+\delta_{i-j,j}} K_{i-j,j} .
\end{equation}
Next we write the probabilities $P(\mathbf{n})$ and $P(\mathbf{n'})$ in the form
\begin{equation}
   P(\mathbf{n}) = \frac{\mathbf{n!}W(\mathbf{n})}{\Omega_{M,N}},\quad
   P(\mathbf{n'}) = \frac{\mathbf{n'!}W(\mathbf{n})}{\Omega_{M,N+1}},\quad
\end{equation}
and note that the stoichiometry of aggregation is such that 
\begin{equation}
   \frac{\mathbf{n'!}}{\mathbf{n!}}
   =
   (N+1)\frac{n_i}{n'_{i-j}(n'_j-\delta_{i-j,j})}
   .
\end{equation}
We substitute into (\ref{app:master:1}) and write the result in the form
\begin{equation}
   \frac{W(\mathbf{n})}{\Omega_{M,N}}
   =
   \frac{2}{N \ens{K}_{M,N+1}}
   \sum_{i=2}^\infty
   \sum_{j=1}^{i-1}
   n_i
   K_{i-j,j} 
   \frac{ W(\mathbf{n'})}{\Omega_{M,N+1}}
\end{equation}
and note the summations now involve only distribution $\mathbf{n}$. Equation (\ref{recursion_Omega:W}) in the paper follows directly from this result. 
 
\section{Concavity of $\Omega_{M,N}$}

\noindent First we note the concave/homogeneous inequality.

\begin{quote}\small

If $f(x,y)$ is concave then  
\begin{equation*}
   f(\alpha x_1 +(1-\alpha)x_2, \alpha y_1 +(1-\alpha)y_2) 
   \geq
   \alpha f(x_1,y_1) + (1-\alpha) f(x_2,y_2) .
\end{equation*}
This is the concave inequality. If $f(x,y))$ is also homogeneous with degree 1, then the right-hand side of the inequality is
\begin{equation*}
    \alpha f(x_1,y_1) + (1-\alpha) f(x_2,y_2) 
    =
    f(\alpha x_1, \alpha y_1) + f((1-\alpha) x_1, (1-\alpha) y_1) 
\end{equation*}
Setting $X_1 = \alpha x_1$, $X_2=(1-\alpha) x_2$, $Y_1=\alpha y_1$, $Y_2=(1-\alpha) y_2$
the concave inequality becomes
\begin{equation*}
   f(X_1+X_2, Y_1+Y_2) \geq f(X_1,Y_1) + f(X_2+Y_2)
\end{equation*}
This is the concave/homogeneous inequality and is satisfied if $f$ is concave and homogeneous with degree 1. Conversely, if this inequality is satisfied and $f$ is homogeneous with degree 1, then $f$ is also concave. This result is obtained by tracing the derivation backwards.  
\end{quote}

\noindent To prove the concavity of $\log\Omega$, write distribution $\mathbf{n}$ of the $(M,N)$ ensemble as a sum of two distributions:
\begin{equation}
   \mathbf{n} = \mathbf{n}_A + \mathbf{n}_B
\end{equation}
such that $\mathbf{n}_A$ belongs in ensemble $(M_A,N_A)$, $\mathbf{n}_B$ belongs in $(M_B,N_B)$, and 
\begin{equation}
   M_A + M_B = M,\quad
   N_A+N_B = N .
\end{equation}
Let $\mathbf{n}^*_A$ be the \mpd\ of the ensemble $(M_A,N_A)$ and $\mathbf{n}^*_B$ the \mpd\ of $(M_B,N_B)$. Their sum $\mathbf{n}^*_A+\mathbf{n}^*_B$ is a distribution in $(M,N)$ and thus we have
\begin{equation}
   \log\Omega_{M_A+M_B,N_A+N_B}
   \geq
   H(\mathbf{n}^*_A + \mathbf{n}^*_B) + \log W(\mathbf{n}^*_A+\mathbf{n}^*_B) .
\end{equation}
The functional on the right-hand side is concave in $\mathbf{n}$ and homogeneous with degree 1, therefore it satisfies the concave/homogeneous inequality:%
\begin{multline}
   \log\Omega_{M_A+M_B,N_A+N_B}
   \geq
   H(\mathbf{n}^*_A + \mathbf{n}^*_B) + \log W(\mathbf{n}^*_A+\mathbf{n}^*_B) 
   \geq
   \\
   H(\mathbf{n}^*_A) + \log W(\mathbf{n}^*_A) + 
   H(\mathbf{n}^*_B) + \log W(\mathbf{n}^*_B)
   = 
   \\
   \Omega_{M_A,N_A}+\Omega_{M_B,N_B} .
\end{multline}
Therefore,
\begin{equation}
   \log\Omega_{M_A+M_B,N_A+N_B} \geq \Omega_{M_A,N_A}+\Omega_{M_B,N_B}
\end{equation}
This is the concave/homogeneous inequality applied to the partition function. Since $\log\Omega_{M,N}$ is homogeneous with degree 1 in $M$ and $N$ the inequality implies that it is concave function of its arguments. It also follows that the intensive partition function $\\log\omega(\xav)=(\log\Omega_M,N)/N$ is concave function of $\xav$. Then
\begin{equation}
\label{app:beta:stability}
   \frac{d^2\log\omega}{d\xav^2} = \frac{d\beta}{d\xav} \leq 0 .
\end{equation}
From Eq.\ (\ref{xav:logq:beta}) we also have
\begin{equation}
   \frac{d\beta}{d\xav} = -\frac{1}{\xav} \frac{d\log q}{d\xav} .
\end{equation}
Combining with (\ref{app:beta:stability}) we obtain
\begin{equation}
\label{app:q:stability}
   \frac{d\log q}{d\xav} \ge 0 .
\end{equation}
Equations (\ref{app:beta:stability}) and (\ref{app:q:stability}) appear as Eq.\ (\ref{stability:beta:logq}) in the main text. 

\begin{figure}
\begin{center}
\includegraphics[width=3.25in]{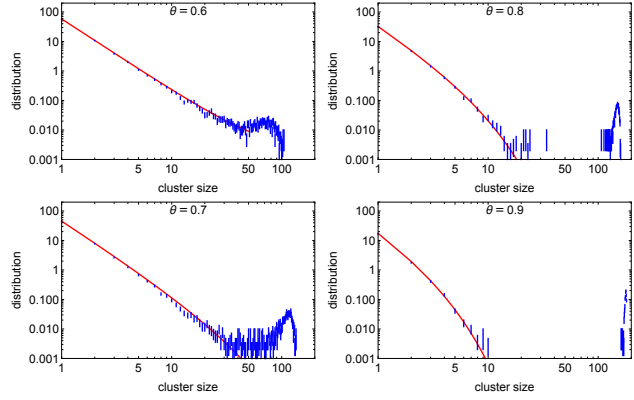}
\end{center}
\caption{The mean distribution in the simulation is calculated as the average of 1000 independent runs. Error bars are estimates of the standard deviation of the calculated average based on the standard deviation of individual runs. }
\label{fig_SM1}
\end{figure}

\section{Monte Carlo Simulations}
The Monte Carlo method was coded in Mathematica using a simulation box with 200 particles. The results shown in the paper are obtained by averaging 1000 independent trajectories, each trajectory representing a list of 200 distributions, starting with $N=200$ and ending with a single cluster. The simulations were rub on a Macbook Pro 3.1 GHz Dual-Core Intel Core i7. The accuracy of the simulation depends entirely on the number of trajectories that are averaged. The calculation of 1000 trajectories takes about 40 min and is a reasonable compromise between computational time and accuracy. 
Accuracy is excellent for small clusters but deteriorates as cluster mass increases, as 
such particles are present at very low concentration and are sampled less frequently  (Fig.\ \ref{fig_SM1}). Even so, the theoretical distribution represents the data very well. The magnitude of the error bars can be decreased by accumulating more trajectories. However, the discrepancies in the precise location of the gel phase are due to the small mass ($M=200$) of the system. Since computational time increases approximately as $M^2$, simulations with larger mass are impractical. We note, however, that the agreement on the location of the gel cluster between theory and simulation (see Fig. 6 of the paper) improves as the size of the cluster gel increases. This behavior is consistent with the claim that discrepancies are due to small size effects. 

\begin{figure}
\begin{center}
\includegraphics[width=3.25in]{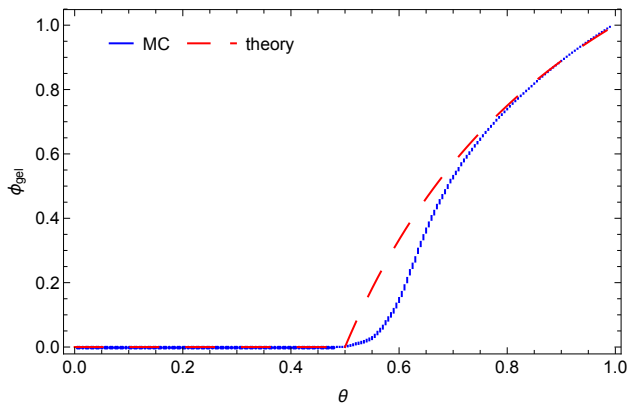}
\end{center}
\caption{The error in $\phi_\text{gel}$ is generally less than 1\% for $\phi_\text{gel}>0.6$.  }
\label{fig_SM2}
\end{figure}

The error in $\phi_\text{gel}$ is very small, less that 1\% over most of the post-gel region (Fig.\ \ref{fig_SM2}).  As a mean property of the distribution, $\phi_\text{gel}$ is calculated very accurately when averaged over 1000 trajectories. The same is true for $\xav_\sol$ (not shown). 


\end{document}